\documentclass[traditabstract]{aa}

\usepackage{graphicx}
\usepackage{txfonts}
\usepackage[colorlinks=true, urlcolor=blue, linkcolor=blue, citecolor=blue, backref=page]{hyperref}
\usepackage{comment}

\usepackage{subcaption}

\usepackage[capitalise]{cleveref}
\crefname{section}{Sect.}{Sects.} 
\crefname{appendix}{Appendix}{Appendices}
\crefname{table}{Tab.}{Tabs.}

\usepackage[colorinlistoftodos]{todonotes}
\newcommand{\kms}{$\rm km\,s^{-1}$}
\newcommand{\degrees}{$^{\circ}$}
\newcommand{\Halpha}{\mathrm{H\alpha}}

\newcommand{\HI}{H\,\textsc{i}}

\usepackage{enumitem}
\usepackage{xcolor}
\newcommand{\referee}[1]{#1}

\begin{document}

   \title{ALMA CO-CAVITY I. Resolved Molecular Gas in Void Galaxies}


    \author{D. Espada \inst{\ref{inst:ugr},\ref{inst:ic1}}
    \and
        S. B. De Daniloff \inst{\ref{inst:iram},\ref{inst:ugr}}
    \and
        S. Duarte Puertas\inst{\ref{inst:ugr}}
    \and
        M. Argudo-Fern\'andez \inst{\ref{inst:ugr},\ref{inst:ic1}}
    \and
        U. Lisenfeld \inst{\ref{inst:ugr},\ref{inst:ic1}}    
     \and 
        S. Verley \inst{\ref{inst:ugr},\ref{inst:ic1}}
    \and 
        I. P\'erez \inst{\ref{inst:ugr},\ref{inst:ic1}}
    \and
        M. I. Rodr\'iguez\inst{\ref{inst:iram}}
    \and 
        T. Ruiz-Lara\inst{\ref{inst:ugr},\ref{inst:ic1}}
    \and 
        R. Garc\'ia-Benito\inst{\ref{inst:iaa}}
    \and
        L. S\'anchez-Menguiano \inst{\ref{inst:ugr},\ref{inst:ic1}}
    \and 
        M. S\'anchez-Portal\inst{\ref{inst:iram},\ref{inst:AsocAstro}}
    \and
        A. Bongiovanni\inst{\ref{inst:iram},\ref{inst:AsocAstro}}
    \and
        S. F. Sánchez \inst{\ref{inst:unam}}
    \and       
        A. Jim\'enez \inst{\ref{inst:ugr}}
    \and
        R. E. Miura \inst{\ref{inst:ugr}}
    \and 
        G. Torres-R\'ios \inst{\ref{inst:ugr}}
    \and 
        P. Villalba-Gonz\'alez \inst{\ref{inst:ubc}}
    \and
        Y. K. González-Koda \inst{\ref{inst:ugr}}
    \and 
        B. Bidaran \inst{\ref{inst:ugr}}
    \and
        M. Alc\'azar-Laynez \inst{\ref{inst:ugr}}
    \and
        A. Zurita \inst{\ref{inst:ugr},\ref{inst:ic1}}
    \and
        E. Florido \inst{\ref{inst:ugr},\ref{inst:ic1}}
    \and
        P. V\'asquez-Bustos \inst{\ref{inst:ugr}}
       }
            
    \institute{Dpto. de F\'isica Te\'orica y del Cosmos, Edificio Mecenas, Campus de Fuentenueva, Universidad de Granada, E-18071, Granada, Spain \label{inst:ugr} {\href{mailto:despada@ugr.es}{\email{despada@ugr.es}}
        \and
               Instituto Carlos I de F\'isica Te\'orica y Computacional, Universidad de Granada, E18071, Granada, Spain \label{inst:ic1}
        \and
               Institut de Radioastronomie Millim\'etrique (IRAM), Av. Divina Pastora 7, N\'ucleo Central 18012, Granada,
            Spain \label{inst:iram}
               }
        \and 
            Instituto de Astrof\'isica de Andaluc\'ia - CSIC, Glorieta de la Astronom\'ia s.n., 18008 Granada, Spain  \label{inst:iaa}
        \and  
            Asociación Astrofísica para la Promoción de la Investigación, Instrumentación y su Desarrollo, ASPID, 38205 La Laguna, Tenerife, Spain \label{inst:AsocAstro}
        \and 
            Universidad Nacional Aut\'onoma de M\'exico, Instituto de Astronom\'ia, AP 106, Ensenada 22800, BC, M\'exico \label{inst:unam}
        \and 
            Department of Physics and Astronomy, University of British Columbia, Vancouver, BC V6T 1Z1, Canada
            \label{inst:ubc}
              }

   \date{}

 
  \abstract
   {The environment plays a key role in galaxy evolution, yet it remains unclear how detailed molecular gas properties and their connection to star formation and stellar content are influenced by both large-scale and local environments. 
   Here we introduce the ALMA CO-CAVITY project, the first interferometric CO(1--0) survey of a large sample of 41 void galaxies (VGs) to characterise in detail their molecular gas properties. 
   It is built over the CAVITY project, offering optical integral field unit (IFU) data
   , enabling a direct, pixel-to-pixel comparison between molecular gas (from ALMA), star formation, and stellar properties, as well as the derivation of their scaling relations. 
   In this work we present ALMA data products for our sample, containing data cubes, moment maps and position-velocity diagrams at angular resolutions of 1\arcsec . We also present molecular gas, stellar mass, and star formation rate surface density maps at a common resolution of 2\farcs5.
   We contextualise our sample against representative unresolved and resolved surveys. 
   While our sample provides a good representation of the VG population and follows the distribution of key properties seen in star-forming galaxy samples, galaxies included in resolved studies from the literature tend to be more massive, less isolated, and located in denser large-scale environments.
   We present global scaling relations for the ALMA CO-CAVITY sample and find that the molecular gas main sequence exhibits the smallest scatter (0.21\,dex), followed by the Schmidt–Kennicutt relation and the star-forming main sequence. From integrated properties alone, we find that these scaling relations for VGs are compatible with those for denser environments. This paper lays the foundation for forthcoming studies exploiting this unique dataset.
   }


   \maketitle
%

\section{Introduction}

It is well established that both local and large-scale environments strongly affect galaxy evolution \citep[e.g.][]{Blanton2009}. The large-scale structure (LSS) of the Universe is probably one of the big successes of the current standard cosmology paradigm, the $\Lambda$--Cold Dark Matter ($\Lambda$CDM) \citep{Peebles1980,Springel2006}. We now know that galaxies tend to cluster together in interconnected structures 
characterised by filaments, walls, dense clusters, and under-dense regions called voids. Gravitational interactions and other physical processes, as well as the availability of gas along cosmic filaments, are key to understanding galaxy evolution \citep{Bond1996}.

Voids are an integral part of the cosmic web \citep[see e.g.][]{van_de_Weygaert2011}, representing around 70\% of the volume of the universe \citep{Pan2012,Cautun2014}. 
Voids originate from the primordial Gaussian field of density fluctuations. As a result of their under density, they represent regions of weaker gravity, and therefore they may expand faster than the Hubble flow. The pristine environment of voids represents an ideal setting for the study of galaxy formation and evolution since they may still be accreting gas from the intergalactic medium (IGM) and are unaffected by the processes modifying galaxies in high-density environments. Void galaxies (VGs), as they evolve in a low density universe with shorter merger histories \citep{Lackner2012} and longer star formation histories (SFHs, \citealt{Dominguez_Gomez2023}), \referee{provide clues on how galaxies grow their stellar mass over cosmic time, which can help us understand what internal mechanisms are important for galaxy evolution, and how these processes depend on the LSS}. 

There is still limited observational information on the properties of VGs and there is no consensus on potential differences to their non-void counterparts. Although some studies indicate that VGs tend to be bluer, of later-type morphologies, and seem to have higher specific star formation rates (sSFR = SFR / $M_\star$) than galaxies in denser environments \citep[e.g.][]{Ricciardelli2014}
, others claim that this increase in star formation (SF) is only marginal \citep{Beygu2016} or not existing \citep{Dominguez_Gomez2022}. \referee{Star formation can be enhanced either by interactions with nearby galaxies \citep[e.g.][]{2013AJ....145..120B} or by gas accretion from cosmic web filaments, which feeds and cools galactic disks, increasing their gas density \citep[e.g.][]{2019MNRAS.482.3403E,2017MNRAS.466.4692K}}. There are also discrepancies regarding the present gas-phase metallicity 
in VGs: some studies show systematic lower metallicities \citep{Pustilnik2011}
, while others show no deviation from the standard mass-metallicity relation \citep{Kreckel2015}. At any rate, it is becoming clearer that the environment has an influence on the stellar mass ($M_\star$) of galaxies, with VGs not following the morphology-density relation \citep{Dressler1980} and preferentially having lower-mass galaxies and more compact than in denser environments \citep{Perez2025}. 
Although the view of how and when SF occurs as a function of the LSS \citep[][]{Dominguez_Gomez2023} and the separation of local and large-scale environmental effects \citep[][]{TorresRios2024} have been studied, a holistic picture that includes the availability of neutral gas in forming stars is still missing.

Data on the gas content (both atomic and molecular) are necessary to understand the mass assembly and the process of SF in these galaxies. 
In addition to being bluer and more star-forming, galaxies tend to be more gas-rich, at least from atomic gas information, than their non-void counterparts, even at fixed mass and
morphology \citep{Florez2021}.
Nevertheless, the resolved properties of the neutral gas content in VGs remain largely unknown. The largest \HI\ survey of VGs to date is based on a sample of 60 objects in the Void Galaxy Survey (VGS, \citealt{Kreckel2012}) sample, where \HI\ was detected for 75\% of the objects and revealed that VGs are gas-rich (masses of about 10$^8$\,M$_\odot$) and many have kinematic and morphological signs of ongoing gas accretion, which may suggest that this population is still in the process of assembling gas, as predicted by simulations. 

Much less is known regarding the molecular gas content and whether its relation to SF holds as a function of the large-scale environments.
\referee{Early single-dish CO observations \citep{1997AJ....114.1753S} and subsequent analyses of VGs 
\citep{2013AJ....145..120B,2015ApJ...815...40D} have shown that galaxies in low-density environments can retain significant molecular gas reservoirs and sustain ongoing star formation, often following similar scaling relations to those in field galaxies (although likely in denser regions). A detailed interferometric CO study of the VG CG-910 by \citet{2026A&A...706A.265S} found an asymmetric molecular gas distribution and an  unusually large (molecular) gas depletion timescale ($t_{\mathrm{dep}}$ $\sim$ 36\,Gyr), indicative of slow evolutionary processes that might be typical of void environments.}
To our knowledge, the largest VG sample studied with molecular gas data is that of \citet{Rodriguez2024}, where single-dish data for 106 VGs are presented and complemented with data from the xCOLD GASS \citep{Saintonge2017} sample for galaxies that matched the same criteria of VG, totalling 200 VGs. No major differences were found by \citet{Rodriguez2024} between VGs and a comparison sample containing galaxies of filaments/walls in the SF efficiency (SFE = SFR/$M_{\rm H_2}$, or star formation rate per unit of molecular gas mass), and in the molecular gas fraction as a function of $M_\star$ for galaxies on the star-forming main sequence (SFMS). 
The lack of differences in molecular gas, SFE, and gas fraction is intriguing, given the observed variations in other global properties, as well as predictions from numerical simulations \citep{2022MNRAS.517..712R,2024ApJ...962...58C}. Nevertheless, such differences may emerge when the properties of galaxies are examined in a spatially resolved manner, instead of as averages over their whole discs.

Recent studies such as ALMaQUEST (66 galaxies, \citealt{Lin2020, Ellison2024}), EDGE-CALIFA survey (125, \citealt{Bolatto2017, Wong2024}), and the PHANGS-MUSE (19, \citealt{Emsellem2022}) surveys, use as strategy to combine the power of optical Integral Field Unit (IFU) data and high resolution CO maps to address how $M_\star$, SFR, metallicity, and other properties derived from IFU data relate to molecular gas. Each of these projects offer complementary strengths focusing mostly on nearby star-forming disc galaxies, albeit with objects in the starburst and transition zones, and typically achieved resolutions better than 1--2\,kpc. In the case of PHANGS-MUSE, better angular resolutions of $\sim$1\arcsec\ (physical scale of $<$ 100\,pc) are targeted, at the cost of a more modest sample size and more local galaxies $D< 20\ \mathrm{Mpc}$. 
Despite the great success of these projects, the roles of local and large-scale environments have not been properly accounted for. While these surveys excelled in studying the physical processes that govern SF at (sub-)kiloparsec scales, as well as their scaling relations, and the potential causes for the deviations from these, they lacked an understanding of the relation with the environment (except in cases of obvious mergers, or deemed not to be obviously interacting). This environmental
context is essential to determine whether the processes governing
molecular gas collapse and SF regulation are
environment-dependent. Investigating the LSS of nearby galaxies is more difficult, as it requires homogeneous spectroscopic information that is often unavailable for objects outside the footprint of a survey such as that of the Sloan Digital Sky Survey (SDSS, \citealt{2011AJ....142...72E}). Moreover, the studied galaxies very unlikely inhabit the most under-dense regions of the cosmic web, making VGs uniquely important for testing the universality of resolved scaling relations. Another caveat from previous high resolution molecular gas studies is that most of the objects are in the high-end of stellar mass and metallicity, partly because they are easier to be detected in CO emission. 

In the ALMA CO-CAVITY project we study a rigorously selected, representative sample of VGs with a unique set of ancillary data, to measure for the first time their molecular gas content, distribution, and kinematics, and relate it to spatially resolved SFRs and SFHs as well as characteristic scaling relations at scales of about 1\,kpc. We then compare with previous surveys which in general represent galaxies in higher density regions and perhaps a greater chance of local interactions. 
The main objectives of the project are further motivated in \cref{sec:motivation}.
In this paper, we begin by describing the sample selection, including galaxy classifications and main properties (\cref{sec:sample}). \referee{Resolved observations and data processing are presented in \cref{sec:observations}, focusing on those conducted with the Atacama Large Millimeter/submillimeter Array (ALMA) and its data reduction  (\cref{sec:alma}). 
In addition, IFU data from CAHA PMAS/PPak and MaNGA are presented in \cref{sec:ifu}. }\referee{Our main results are presented in \cref{sec:results}. We obtain $M_\mathrm{H_2}$ (\cref{subsec:molecularmass}), ALMA CO-CAVITY data products (channel maps, moment maps, and position-velocity diagrams) at native angular resolution of about 1\arcsec (\cref{subsec:atlas}), as well as  molecular gas ($\Sigma_\mathrm{H_2}$), SFR ($\Sigma_\mathrm{SFR}$), and stellar ($\Sigma_\star$) surface densities at a common angular resolution of 2\farcs5 (\cref{subsec:resolved_physical_properties}). In \cref{sec:scaling_relations} we present the main global scaling relations between these three parameters: the Schmidt-Kennicutt law (SK), the molecular gas main sequence (MGMS), and the SFMS. 
We discuss our results in \cref{sec:discussion}  and present our conclusions in \cref{sec:conclusions}. }

Throughout this paper we adopt a $\Lambda$CDM cosmology, $H_0$ = 70 \kms Mpc$^{-1}$, $\Omega_m$ = 0.30, and $\Omega_{\Lambda}$ = 0.70. As initial mass function (IMF), we chose \citet{Chabrier2003}, and modifications to external data have been performed for comparison. 
We note that previous resolved scaling relations studies have usually used a Salpeter IMF \citep{Salpeter1955}, partly because it is the default option adopted by the processing package for IFU data, \texttt{PIPE3D} \citep{Sanchez2016a}. The conversion from Salpeter to Chabrier IMFs is achieved by multiplying the former by a factor 0.61 for $M_\star$ and 0.63 for SFR, and from Kroupa \citep{Kroupa2001} 
to Chabrier IMFs a factor 0.940 and 0.924, respectively \citep{Madau_Dickinson2014}.

\section{Motivation of the ALMA CO-CAVITY project}
\label{sec:motivation}

\referee{ALMA CO-CAVITY extends the objectives of Calar Alto Void Integral-field Treasury surveY\footnote{\url{https://cavity.caha.es}} (CAVITY, \citealt{Perez2024})} to deliver spatially resolved molecular maps for a subset of VGs. Combined with the wealth of information from IFU data, it allows to study, particularly for this kind of galaxies, the link between molecular gas, SF and the stellar properties. Besides the sample being composed of VGs, complementing previous studies focusing on denser and more diverse cosmic environments, this project aims to extend the parameter space of previous resolved studies towards lower $M_\star$ galaxies, down to 10$^9$\,M$_\odot$, to provide a more complete picture of molecular gas regulation in this regime and better understand galaxy evolution. 
The selection criteria of the ALMA CO-CAVITY sample are detailed in \cref{sec:sample}. 
The main scientific drivers of ALMA CO-CAVITY are: 

\begin{itemize}[leftmargin=0pt, itemsep=0pt, topsep=0pt, label={}]

\item ~~~1) Bridging unresolved and resolved molecular gas studies:
Existing unresolved studies (e.g. IRAM 30m CO-CAVITY, xCOLD GASS) across different environments provide global molecular gas properties, but they lack spatially resolved information on how gas is distributed within galaxies. A question then arises of whether the resolved molecular gas (fraction) depends on density environment. ALMA CO(1--0) observations,  combined with optical IFU data, enable to bridge this gap for specifically selected VGs by linking resolved gas to SF regions at 1\,kpc scales.

\item ~~~2) Extending previous projects with resolved data in low density regions and low-mass galaxies: Previous projects such as ALMaQUEST and EDGE-CALIFA have provided crucial insights into the resolved molecular gas properties of galaxies, but they usually target nearby massive galaxies located in relatively dense environments.  
On the other hand, low- to intermediate-mass galaxies (below $\sim 10^{10}$\,M$_\odot$) that populate the low-density regions of voids may follow different accretion, SF and feedback mechanisms. The resolved Schmidt-Kennicutt law (rSK), the resolved molecular gas main sequence (rMGMS), and the resolved SF main sequence (rSFMS) have been well-studied for massive galaxies, but low-mass galaxies in voids may follow different trends. In addition, it may reveal a stronger influence of gas-phase metallicity on the CO-to-H$_2$ conversion factor ($\alpha{_\mathrm{CO}}$) and reveal a new region in the resolved fundamental metallicity plane parameter space.

\item ~~~3) Probing the fundamentality of scaling relations in the extreme low-density environment regime:
A question that has recently attracted attention is which scaling relation is the most fundamental. \citet{2019ApJ...884L..33L} suggested that the rSFMS is not fundamental, but rather a result of combining the rMGMS and rSK. Using the ALMaQUEST sample, no statistical evidence for a relation between $\Sigma_\star$ and $\Sigma_{\rm SFR}$ once the rMGMS and rSK are accounted for \citep{2022MNRAS.510.3622B}. Extending these analyses to the extreme low-density regime of voids provides a stringent test of whether such relations are truly universal and how do external influences (e.g. mergers, cosmic web flows) may impact. 

\item ~~~4) Probing environmental effects on local (1~kpc) scales: Previous high resolution studies do not consider the potential role of LSS on local (galactic) scale properties. 
 The rSK, rSFMS, and rMGMS relations show significant galaxy-to-galaxy variations in intercept, slope, and scatter, indicating that none of these relations is truly universal 
 \citep[e.g.][]{2021MNRAS.501.4777E}. Within a single galaxy there are also systematic variations that depend on local environment, galaxy structure, and SFE \citep[e.g.][]{2020MNRAS.493L..39E}. A measure of local dynamical equilibrium pressure has also been proposed as a more fundamental indicator for predicting $\Sigma_{\rm SFR}$ than $\Sigma_{\rm H_2}$ and $\Sigma_{\star}$ \citep{2021MNRAS.503.3643B,Ellison2024}.   
 It is essential to probe whether properties such as the SFE as a function of radius change systematically with LSS. Radial analyses from ALMaQUEST demonstrate that quenching proceeds
  through a competition between gas availability and SFE as a function
  of radius \citep{2024ApJ...964..120P}.  
 Whether such mechanisms operate similarly in galaxies in the most under-dense regions
  remains an open question that CO-CAVITY is uniquely positioned to
  address.    

\end{itemize}

\section{Sample}\label{sec:sample}

\referee{ The galaxies used in this work are drawn from the CAVITY sample \citep{Perez2024}, which is based on the Catalogue of Cosmic Voids \citep{Pan2012}.}
 \citet{Perez2024} selected galaxies in the redshift range 0.005\,$<$\,$z$$<$\,0.050, with inclinations between 20\degrees and 70\degrees, requiring at least 20 galaxies per void and excluding objects near the edges of the SDSS footprint. A total of 15 voids were chosen to adequately sample the void effective radius ($R_{\rm eff}$), considering volume number densities,  
 as well as the right ascension in order to maximize observability from CAHA.  
 Objects belonging to clusters that overlapped with the \citet{Tempel2017}'s catalogue were removed. A visual inspection followed, discarding galaxies with very bright stars within the PMAS field of view and objects with too low surface brightness.  
 From the resulting sample, 300 were selected to uniformly cover the parameter space of the color–magnitude diagram \citep{Perez2024}.

From the CAVITY sample, the selection criteria that yields the 41 galaxies in ALMA CO-CAVITY are:
1) Declination $\rm Dec < 30$\degrees, so that they are easily observable from the ALMA site,
2) $M_{\star}>10^9\ {\rm M_\odot}$,
3) metallicity $\rm 12+\log(O/H) > 8.4$ (otherwise, the galactic CO-to-molecular gas mass ratio may not apply, \citealt{Bolatto2013}),
4) with IFU data, either from CAVITY, or archival data (MaNGA, CALIFA, VGS),
5) with \HI\ data from the literature, and 
6) with optical sizes of at least $d_{25} > 20\arcsec$ (where $d_{25}$ = 3 $\times$ $r_{90}$, being $r_{90}$ the radius containing 90\% of the Petrosian flux in the $r$-band).
Finally, we prioritised galaxies that belonged to a few voids in order to have information as complete as possible for individual voids, and to minimise time spent on calibration as most members of a void are grouped 
to be observed subsequently and share calibration data.

\begin{figure*}
    \centering
    \includegraphics[width=\textwidth]{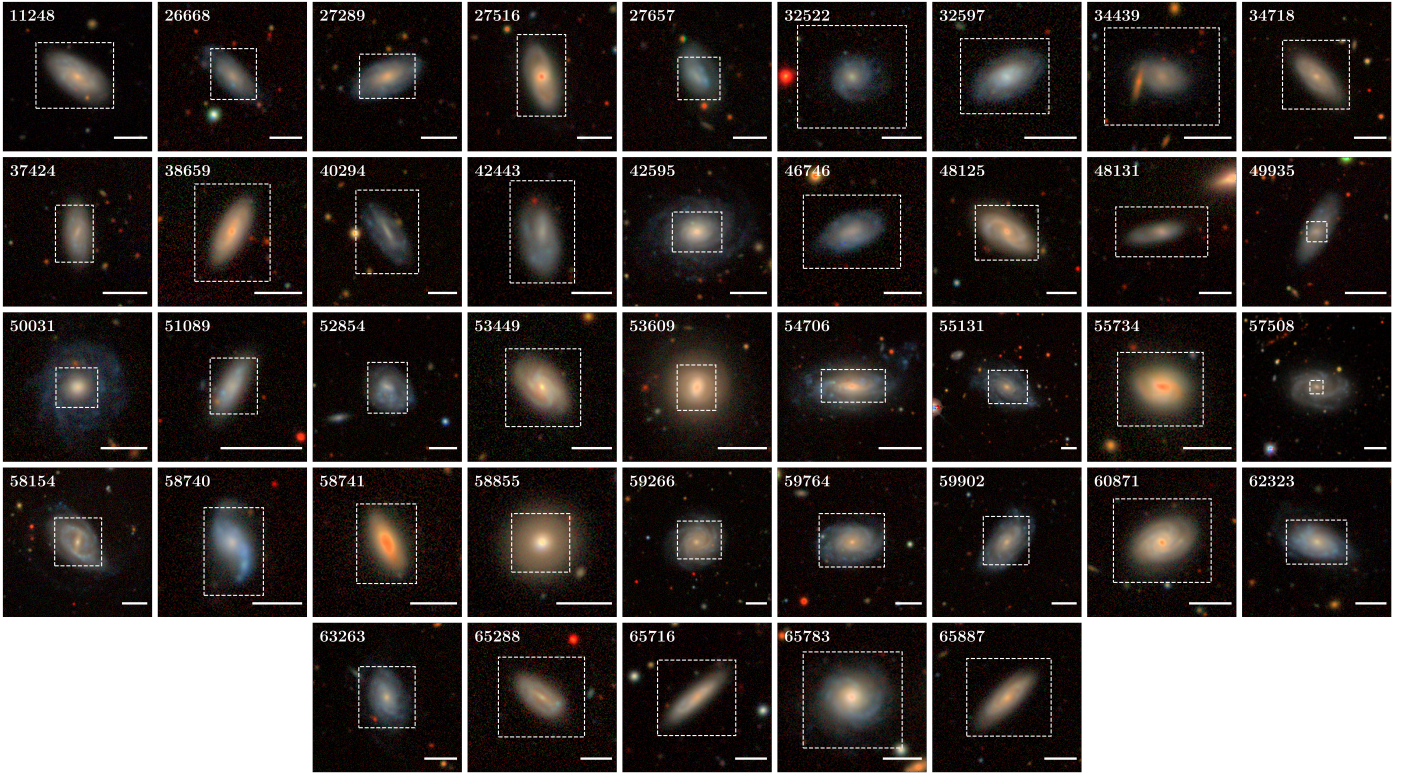}
    \caption[mugshot]{
    False RGB colour images (red - $z$ band, green - $r$ band, blue - $g$ band) from DECaLS legacy survey of the 41 ALMA CO-CAVITY VGs. In each panel, we show the CAVITY ID of the galaxy in the upper-left corner, and a scale bar of 10\,kpc in the lower-right corner. The dashed rectangle corresponds to the field shown in \cref{fig:mugshot_molecular_all_galaxies} for the CO(1--0) moment-0 maps.}
    \label{fig:mugshot_all_galaxies}
\end{figure*}

In \cref{fig:mugshot_all_galaxies} we present the optical images of the 41 galaxies composing the ALMA CO-CAVITY sample. The galaxies are mostly blue spiral galaxies without strong signs of interaction.
In \cref{tab:galaxy_properties} we also list the individual main properties for each galaxy, mainly from SDSS DR7, the MPA/JHU catalogue \citep{Kauffmann2003a, Brinchmann2004, Tremonti2004, Salim2007} and \citet{Dominguez_Sanchez2018}. See Sect. 3.2 of CO-CAVITY \citep{Rodriguez2024} for more details on how the $M_\star$ and SFRs were obtained.
This set of 41 VGs ensures wide coverage in terms of 
stellar mass ($\log\left(M_{\star}/{\rm M_{\odot}}\right) = [9.1,\, 10.7]$), SFR ($\rm \log\left(SFR/\left[M_{\odot}\,yr^{-1}\right]\right) = [-1.0,\, 1.0]$), sizes ($d_{25}$ = [9, 46]\,kpc),
morphological types (T-Type = [-2.3, 5.7]), and 
distances to the void’s centres ($r / R_{\rm eff}$  = [0.2, 0.8], where $R_{\rm eff}$ is the effective radius of the void). 

\begin{figure}
    \centering
    \includegraphics[width=0.8\columnwidth]{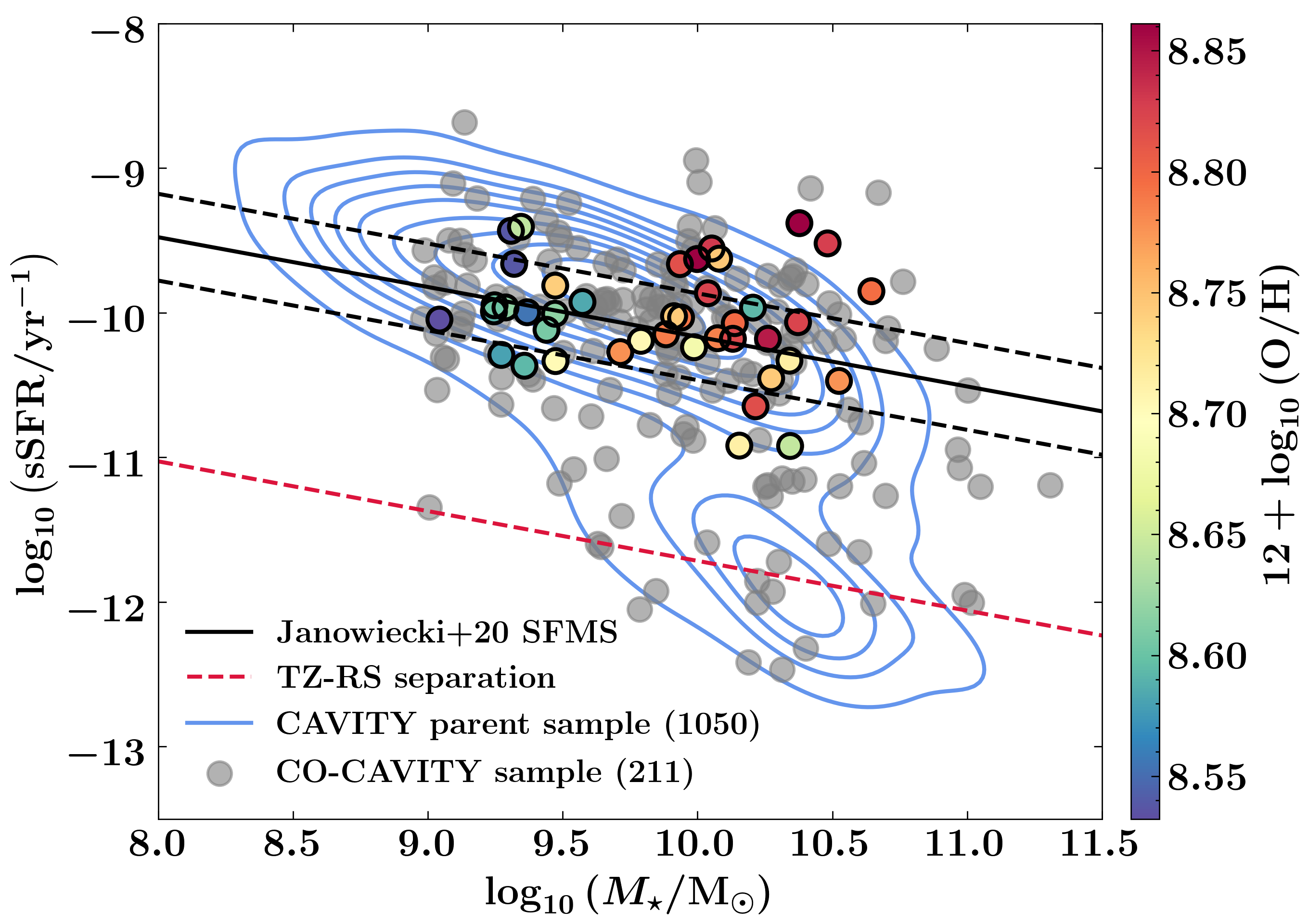}
    \caption{
    ${\rm sSFR}-M_{\star}$ diagram for the 41 galaxies in the ALMA CO-CAVITY sample, compared to the galaxies from the  CAVITY parent sample \citep{Perez2024} with SFR and $M_\star$ estimated (blue contours), and the CO-CAVITY VGs (grey markers). The colour bar denote gas metallicity $\rm 12+\log(O/H)$. We also show the \cite{Janowiecki2020}'s SFMS relation (black curve) used by \cite{Rodriguez2024} to classify the CO-CAVITY sample according to their SF activity, and the separation between the transition zone (TZ) and the red sequence (RS).}
    \label{fig:sSFR_Mstar}
\end{figure}

In \cref{fig:sSFR_Mstar} we show the sSFR vs $M_{\star}$ diagram for the ALMA CO-CAVITY sample. We present the SFMS extracted from \cite{Janowiecki2020} (their Eq.1) with the $\pm0.3$ dex uncertainty range (approximately 1$\sigma$).  
We adopt the same thresholds to classify galaxies, i.e. $\rm SFMS \pm 0.3$ dex for the intrinsic scatter of the SFMS, 1.55\,dex below the SFMS to separate galaxies in the Transition Zone (TZ) and the Red Sequence (RS) \citep{Janowiecki2020,Rodriguez2024}. According to this SF classification, our galaxies are classified mainly as star-forming (25 objects), although there are 10 SBs and 6 TZs. There are no galaxies in the RS. 

\cite{Rodriguez2024} studied the molecular gas of VGs using (unresolved) IRAM 30m CO(1--0) observations. Out of the VGs in the CO-CAVITY sample, 37 galaxies are within the ALMA CO-CAVITY sample.  
For instance, 5 galaxies did not have a classification in \cite{Rodriguez2024} because they either were not observed as part of the (IRAM 30m) CO-CAVITY (that is the case of galaxies \texttt{27289}, \texttt{55734}, \texttt{57508}, and \texttt{59266}), or it was deemed to be contaminated by AGN using the BPT diagram (galaxy \texttt{53609}). We thus classify  the former four galaxies using the methodology from \cite{Rodriguez2024}. For galaxy \texttt{53609}, using the BPT map (as in \cref{subsec:resolved_physical_properties}) obtained from IFU data, we confirmed that although potentially contaminated as a composite, it is not AGN dominated and therefore decided not to exclude it.  
Considering it as not AGN-dominated, it would have been classified as TZ.

In \cref{tab:galaxy_environments} we present the environmental properties of the ALMA CO-CAVITY sample, including the distance to the centre of the void ($r/R_{\rm eff}$), as a fraction to 
$R_{\rm eff}$, which corresponds to the radius of a spherical void of equal volume to that identified by \cite{Pan2012}, and the number of physically bound neighbours within 0.45\,Mpc projected radius and $\pm$160\,km\,s$^{-1}$ in line-of-sight velocity ($N_{\rm neigh, TR24}$) compiled from \citet{TorresRios2024}, based on the catalogue by \citet{Tempel2017}. Note that if $N_{\rm neigh, TR24}$ = 0 then they are called `singlets', meaning that they do not have any physically bound neighbour. We also list, using the NSA catalogue and considering a volume-limited sample of neighbours to avoid redshift bias, LSS neighbours within 1\,Mpc projected radius and $\pm$500\,km\,s$^{-1}$ ($N_{\rm neigh, AF15}$) following the criteria in \citet{Argudo2015}. If $N_{\rm neigh, AF15}$ = 0 then galaxies are called `isolated'. 
In addition, we present the projected distance to the nearest LSS neighbour within the latter volume ($d_{\rm nn}$), projected local number density $\eta_{\rm k}$ and at the distance to $k^{\rm th}$ nearest neighbour. We also include whether an asymmetry is apparent from visual inspection of the optical images. 

From inspection to the table, one of the most notable difference is that the majority of the galaxies in the ALMA CO-CAVITY sample are isolated. Therefore, in addition to fulfilling one of the project's objectives of minimising large-scale density, we also minimise potential contamination from the local environment. An 83\% (34 objects) of the ALMA CO-CAVITY sample are singlets, and 65\% (27) are isolated.

\subsection{Comparison with other samples}
\begin{figure*}
    \centering
    \includegraphics[width=0.45\textwidth]{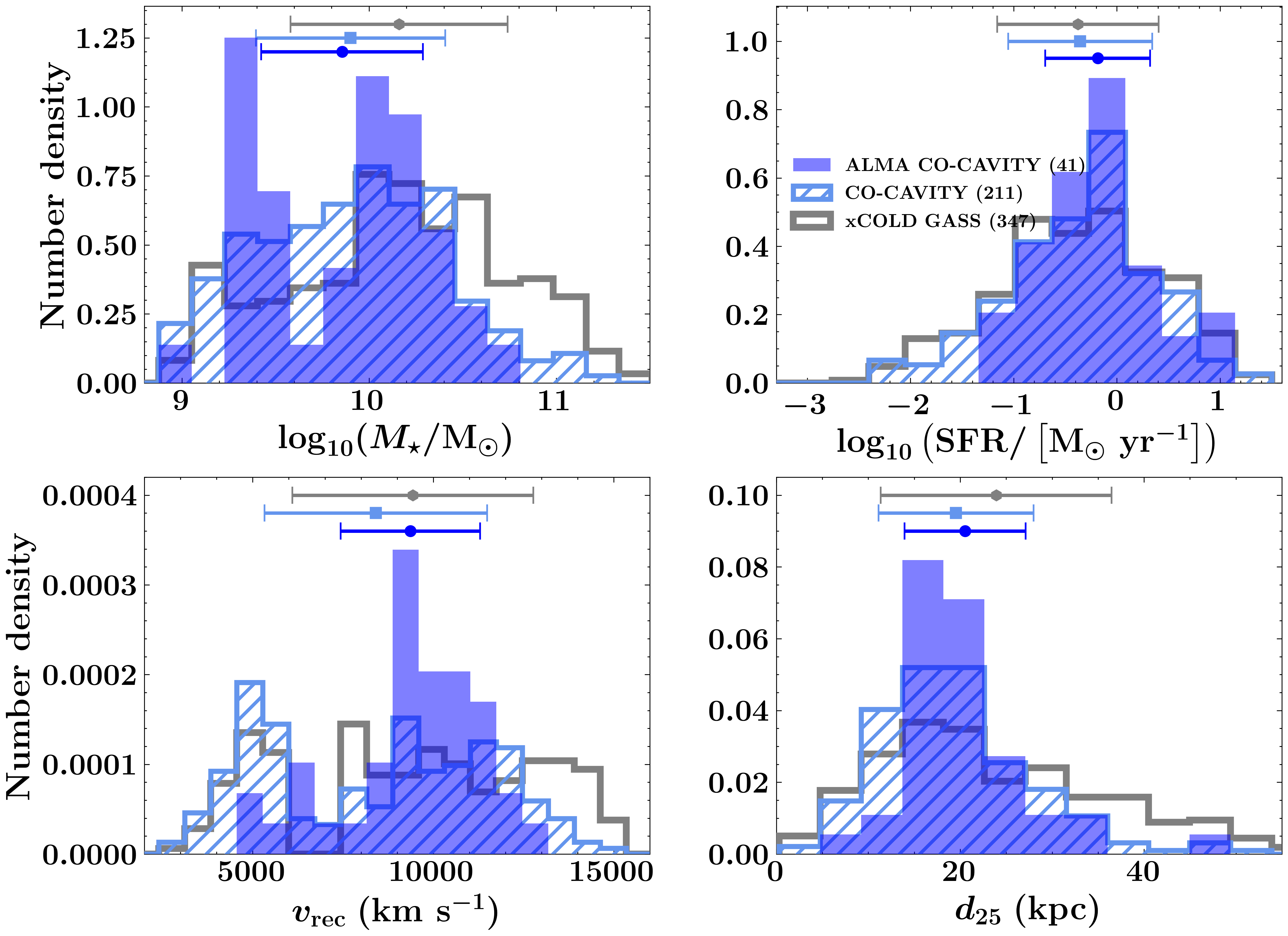}
    \includegraphics[width=0.45\textwidth]{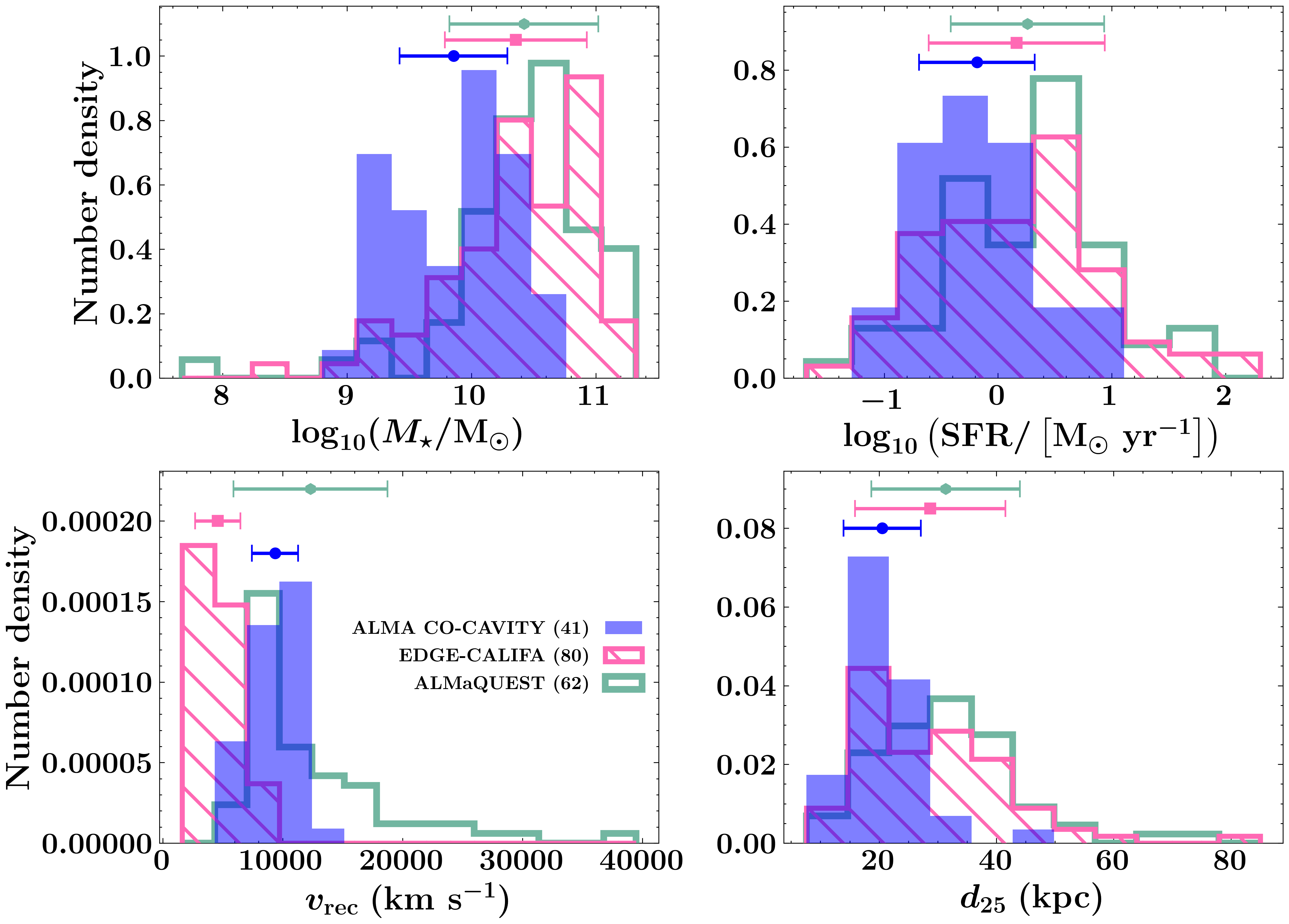}
    \caption{
    Distributions of $M_\star$, $\rm SFR$, $v_{\rm rec}$, and $d_{25}$ for the ALMA CO-CAVITY sample (blue filled histogram), in comparison with (Left panels) unresolved studies: the CO-CAVITY \citep[blue hatched,][]{Rodriguez2024} and xCOLD GASS \citep[dark grey unfilled,][]{Saintonge2017} samples; (Right panels) and resolved studies: the EDGE-CALIFA \citep[magenta hatched,][]{Bolatto2017} and ALMaQUEST \citep[green unfilled,][]{Lin2020} samples. Horizontal bars indicate the mean values and standard deviations.
        }
    \label{fig:histograms_properties}
\end{figure*}

\cref{fig:histograms_properties} shows the distributions of $M_\star$, $\rm SFR$, recession velocity $v_{\rm rec}$, and $d_{25}$ for the ALMA CO-CAVITY galaxies in comparison with CO-CAVITY and xCOLD GASS. For each survey, their properties were obtained following the same methodology as for (ALMA) CO-CAVITY.  
We found counterparts for 347 of the 353 selected xCOLD GASS galaxies (selected regardless of environment) in the MPA/JHU catalogue. We therefore used these galaxies to compare the distribution with that of the ALMA CO-CAVITY sample. 
The ALMA CO-CAVITY sample represents well the probability density distributions of the CO-CAVITY \citep{Rodriguez2024} sample for $M_\star$, SFR, and optical size (\cref{fig:histograms_properties} left). However, objects with SFR below $0.1\ \mathrm{M_{\odot}\ yr^{-1}}$ are missing. A note of caution is that likely as a result of the nature of VGs, the distribution of other samples such as xCOLD GASS is skewed towards more massive galaxies. 

Similarly to \cref{fig:histograms_properties} (left) for the comparison with samples in unresolved studies, on the right panels we present the same comparison of the distribution of the main parameters with two of the main resolved studies, EDGE-CALIFA and ALMaQUEST. Note that 80 of the 125 EDGE-CALIFA galaxies and 62 of the 66 ALMaQUEST galaxies were found in the MPA/JHU catalogue for a comparison of the sample distributions. 
Larger differences arise between the ALMA CO-CAVITY distribution when compared with those of EDGE-CALIFA and ALMaQUEST. We observe that the ALMA CO-CAVITY galaxies extend well below the $M_\star$ and SFR of their distribution, with a greater fraction of galaxies  below $M_\star$ $\sim$ $10^{10}$ M$_\odot$ and SFR  $\sim$ 1\,M$_\odot$\,yr$^{-1}$. The distances (and thus recessional velocities) of ALMA CO-CAVITY galaxies are comparable to those in the ALMaQUEST sample, except the portion of mergers in their sample that forms the wing in the velocity distribution toward large values. On the other hand, the  distances of ALMA CO-CAVITY galaxies are larger than those of EDGE-CALIFA galaxies.  
Finally, the ALMA CO-CAVITY galaxies typically exhibit smaller optical sizes ($d_{25}$ in kpc), with a narrower distribution compared to those in ALMaQUEST and EDGE-CALIFA. 
These properties partly reflect the intrinsic nature of galaxies as inhabitants of void environments, i.e. they are characterised by smaller sizes, $M_\star$, and lower SFR. 
 While ALMaQUEST and EDGE-CALIFA have extensively characterised the distribution of molecular gas in more massive, higher-density environments, ALMA CO-CAVITY provides a natural benchmark for assessing how VGs may differ.
 
\begin{table*}[h!]
    \centering
    \caption{Main galaxy properties of the ALMA CO-CAVITY sample}
    \label{tab:galaxy_properties}
    \resizebox{\hsize}{!}{
    \begin{tabular}{cccccccccccccc}
    \hline\hline
        ID   & R.A.         & Dec        & $z$           & $D_{\rm L}$    & $d_{25}$    & Incl.      & PA        & T-type & $\log(M_\star)$ & $\rm \log(SFR)$ & SF type & 12 + log(O/H) \\
             & (\degrees)   &  (\degrees)&               & (Mpc)          & ($\arcsec$) & (\degrees) & (\degrees)  &       & log([M$_\odot$]) & log([M$_\odot$\,yr$^{-1}$]) & \\
        (1)  & (2)   & (3)  & (4)        & (5)           & (6)            & (7)         & (8)        & (9)              & (10)  & (11) & (12)       & (13) \\ \hline

        11248 & $13^{\mathrm{h}}58^{\mathrm{m}}15.23^{\mathrm{s}}$ & $3^\circ59{}^\prime53.88{}^{\prime\prime}$ & 0.0300 & 131.6 & 35.6 & 58.6 & 55.5 & 4.08 & $10.10 \pm 0.09$ & $-0.103 \pm 0.010$ & SF & 8.8 \\
        26668 & $8^{\mathrm{h}}10^{\mathrm{m}}54.51^{\mathrm{s}}$ & $26^\circ41{}^\prime48.38{}^{\prime\prime}$ & 0.0365 & 160.8 & 29.7 & 62.5 & 44.2 & 4.93 & $9.50 \pm 0.08$ & $-0.536 \pm 0.007$ & SF & 8.6 \\
        27289 & $8^{\mathrm{h}}35^{\mathrm{m}}12.66^{\mathrm{s}}$ & $28^\circ37{}^\prime19.52{}^{\prime\prime}$ & 0.0356 & 156.7 & 27.5 & 58.3 & 112.9 & 4.31 & $9.8 \pm 0.1$ & $-0.41 \pm 0.06$ & SF & 8.7 \\
        27516 & $8^{\mathrm{h}}32^{\mathrm{m}}32.96^{\mathrm{s}}$ & $29^\circ56{}^\prime17.14{}^{\prime\prime}$ & 0.0300 & 131.3 & 34.0 & 49.1 & 11.0 & 2.26 & $10.50 \pm 0.09$ & $0.053 \pm 0.010$ & SF & 8.8 \\
        27657 & $8^{\mathrm{h}}18^{\mathrm{m}}11.51^{\mathrm{s}}$ & $26^\circ39{}^\prime14.13{}^{\prime\prime}$ & 0.0383 & 168.9 & 27.2 & 57.2 & 33.0 & 5.71 & $9.30 \pm 0.06$ & $-0.126 \pm 0.007$ & SB & 8.5 \\
        ...\\
        \hline
    \end{tabular}}

\tablefoot{
(This table is truncated to the first 5 entries. The complete table is presented in its entirety in the online version of the publication.) (1) ID: CAVITY identifier in the CAVITY sample. 
(2) and (3) R.A.: Right Ascension and Dec.: Declination in degrees in the ICRS system from SDSS DR7. 
(4) $z$: redshift from SDSS DR7. 
(5) $D_{\rm L}$: luminosity distance. 
(6) $d_{25}$: diameter at a 25 mag isophote (from the CAVITY \texttt{voids15} master catalogue), obtained from the $r$-band Petrosian radius $R90,r$ from the SDSS as 3$\times R90,r$. 
(7) Incl.: inclination from the CAVITY database \citep{Perez2024,Garcia_Benito2024} (from the \texttt{voids15} master catalogue), defined as arccos($b$/$a$) where $b$ and $a$ are the lengths of the minor and major axes of the galaxy, respectively. 
(8) PA: position angle, from North to East. In the majority of cases we rely on SDSS PA from $r$-band, and complemented the PA from LEDA in cases where SDSS PA was unreliable; if both were deemed unreliable, the PA was corrected by hand based on optical emission.
(9) $\rm T-type$: morphological type from \cite{Dominguez_Sanchez2018}. 
(10) log($M_\star$): decimal logarithm of stellar mass from the MPA-JHU DR7 catalogue \citep{Kauffmann2003a, Salim2007}.
(11) log(SFR): decimal logarithm of H$\alpha$ based Star Formation Rate (SFR) calculated following the method presented in \cite{Duarte_Puertas2017}.
(12) SF type: classification into SF (star forming), TZ (transition zone), SB (starburst) by \citet{Rodriguez2024} following \cite{Janowiecki2020} (SF galaxies within the SFMS ($\pm$ 0.3\,dex), galaxies that lie in the transition zone (TZ), and galaxies below the TZ (sSFR $<$ SFMS-1.55\,dex)). 
(13) 12 + log(O/H): gas-phase metallicity estimated using \cite{Pettini_Pagel2004} [\ion{N}{II}]/H$\alpha$ and [\ion{O}{III}]/H$\beta$ line ratios (O3N2) indicator.}

\end{table*}

 As for the environmental properties, applying the quantification of the LSS environment to ALMaQUEST and EDGE-CALIFA samples in the same way as performed for the ALMA CO-CAVITY sample, based on the NSA catalogue and considering a volume-limited sample of neighbours, we find that only a 20--25\% of the galaxies in these two samples are isolated. 

\begin{table*}[h!]
    \centering
    \caption{Environmental properties of the ALMA CO-CAVITY sample}
    \label{tab:galaxy_environments}
    \begin{tabular}{rrrrrrrrrr}
    \hline\hline
    ID & Void & $r/R_{\rm eff}$ & $N_{\rm neigh,\ TR24}$ & $N_{\rm neigh,\ AF15}$ &  $d_{\rm nn}$ & $\eta_{\rm k}$ & $d_{\rm kn}$ & $\eta_{\rm k,\ corr}$ & Asymmetry \\
     &  &   &  &   & (Mpc)  &   & (Mpc)  &  &\\
    (1) & (2)  & (3)  & (4)  & (5)  & (6)  & (7)  & (8) & (9) & (10)\\ \hline
    
    11248 & 439 & 0.49 & 0 & 0 & - & - & - & - & \\
    26668 & 139 & 0.74 & 0 & 0 & - & - & - & - & \\
    27289 & 139 & 0.41 & 0 & 0 & - & - & - & - & \\
    27516 & 139 & 0.77 & 1 & 1 & 0.15 & - & - & -0.18 & \\
    27657 & 139 & 0.73 & 0 & 2 & 0.16 & -0.29 & 0.77 & 0.00 & Y\\
    ...\\
    \hline
    \end{tabular}

    \tablefoot{
    (This table is truncated to the first 5 entries. The complete table is presented in its entirety in the online version of the publication.)
    (1) ID: the catalogue identifier in the CAVITY sample,
    (2) Void: Void ID to which the galaxy belongs from \citet{Pan2012},
    (3) $r/R_{\rm eff}$: Normalized void-centric distance, expressed as $r/R_{\rm eff}$, where $R_{\rm eff}$ is the effective radius of the void.
    (4) $N_{\rm neigh,\ TR24}$: Number of neighbouring bound galaxies within 0.45\,Mpc and $\pm$160\,km\,s$^{-1}$ in line-of-sight velocity from \citet{TorresRios2024} based on the catalogue by \citet{Tempel2017}. If $N_{\rm neigh,\ TR24} = 0$ then they are called singlets, meaning they do not have any physically bound neighbour.
    (5) $N_{\rm neigh,\ AF15}$: Number of neighbouring galaxies within 1\,Mpc projected radius and $\pm$500\,km\,s$^{-1}$ in line-of-sight velocity, as in \citet{Argudo2015}. If $N_{\rm neigh,\ AF15} = 0$ then they are called isolated.
    (6) $d_{\rm nn}$: Projected distance to the nearest neighbour within this volume in Mpc.
    (7) $\eta_{\rm k}$: Local projected density, estimated using galaxies within 1\,Mpc and $\pm$500 km/s from the NASA-Sloan Atlas (NSA) catalogue. Only galaxies with $M_r < -18.3$ are considered, to ensure completeness up to redshift $z < 0.044$.
    (8) $d_{\rm kn}$: Projected distance to the $k^{\rm th}$ nearest neighbour within this same volume in Mpc.
    (9) $\eta_{\rm k,\ corr}$: Corrected local projected density, estimated without applying an absolute magnitude cut to the galaxy sample, in order to include fainter companions and avoid redshift-dependent selection biases.
    If no galaxies or only one galaxy is found in the volume a '-' sign is shown because the density estimates are not meaningful.
    (10) Asymmetry: Optical asymmetry by visual inspection  of the optical image.}

\end{table*}

\section{Observations and Data}
\label{sec:observations}

\subsection{ALMA Observations and Data reduction}
\label{sec:alma}

The observations were carried out with ALMA during Cycles 9 and 10 using the band 3 receiver through program 2022.1.00482.S (PI: D. Espada). The angular resolution range requested was $1\farcs0-1\farcs5$ ($\sim0.64-0.96$\,kpc). This ensures a resolution superior to that of optical IFU data from CAVITY, which is of $2\farcs5$. The requested largest angular scale (LAS) was $15\arcsec$ ($\sim10\, \rm kpc$). These two conditions required the use of either a single configuration, ALMA main array configuration C$-3$ for 15 galaxies, or a combination of C$-1$ and C$-4$ for 26 galaxies.  
The field of view (Half Power Beam Width --HPBW-- of the primary beam) is $\sim54\arcsec$ at $\sim$115\,GHz, that of a single pointing. The optical sizes of the galaxies are in most cases $d_{25}$ $<$ 60\arcsec, so the CO(1--0) emission can be well observed in a single pointing  with ALMA.  
The absolute flux calibration accuracy of ALMA data in band 3 is $\sim5\%$\footnote{This represents a $2\sigma$ accuracy,  
see ALMA Technical Handbook (\url{https://almascience.eso.org/documents-and-tools/cycle11/alma-technical-handbook}).}. Detailed information about the ALMA observational setup and data processing is in Appendix \ref{appendix:informationObservations}. 

We retrieved data processed with the two versions of the ALMA CASA pipeline, either \texttt{6.4.1.12/2022.2.0.64} or \texttt{6.5.4.9/2023.1.0.124}, as provided by the observatory.
For those galaxies observed with a single array configuration (i.e. C$-3$), we directly used primary beam corrected clean maps at native resolution after inspecting the reliability through Quality Assurance Level 2 (QA2) reports and other data products. Data combination is explained in \cref{subsec:datacombination}.

Detailed information about the final data cubes at native angular resolution can also be found in Appendix \ref{appendix:informationObservations}.
\cref{tab:ALMA_products} shows the final properties of the datacubes for each galaxy, including the RMS, synthesised beam and if they were classified as detections, tentative or non-detections (see \cref{subsec:masks} for details).

\subsubsection{Data combination}
\label{subsec:datacombination}

In the case of the 26 galaxies that were observed with two ALMA configurations, we combined the measurement sets (MS) for each dataset corresponding to compact and extended configurations. During this process, we applied a continuum subtraction of each MS with the {\tt uvcontsub} task and we used the task {\tt concat} to combine the MS files with a frequency tolerance of 30\,MHz. We obtained the clean data cubes using {\tt tclean} with {\tt ROBUST} parameter 0.5, a sensitivity threshold of $\rm 2\times \sigma$, a channel width of 10\,\kms, and a pixel size of 0\farcs1. We used the auto-masking {\tt auto-multithresh} option and applied the primary beam correction to obtain the final clean data cube. 

\subsubsection{Rebinning, smoothing and regridding}

The datacubes were rebinned to a common spectral resolution of 10\,\kms\ using CASA's {\tt imrebin} task. 
In addition to providing datacubes at native angular resolution, we also obtained datacubes smoothed to a circular PSF of 2\farcs5 using {\tt imsmooth} and adopted the same pixel size of either 1\farcs0 for PPak data or 0\farcs5 for MaNGA data, using {\tt imregrid}. This downgrade in resolution was performed to allow direct comparison between ALMA CO-CAVITY and the IFU data products. For more details, see ALMA CO-CAVITY Paper II (De Daniloff et al., in prep.).

\subsubsection{Masks and detections}
\label{subsec:masks}

We obtained a mask for each datacube at the original resolution of the data, and then rebinned spectrally at a resolution of 10\,$\rm km\, s^{-1}$ and smoothed spatially at a resolution of 2\farcs5. We used the {\tt Source Finding Application, SoFiA 2} pipeline \citep{Westmeier2021} to obtain the 3D masks. We first cropped the datacube in a range of $\pm 300$ $\rm km\, s^{-1}$ around the optical velocity to avoid false detections outside the emission region. We used the primary beam datacube as input weights. We used the S+C (Smooth + Clip) finder provided by \texttt{SoFiA 2} to identify the regions with CO emission.  
 It operates by applying iterative spatial and spectral smoothing, followed by clipping at a specified noise threshold to mask (S/N = 3.5). 
The linking module, working as a friends-of-friends algorithm, was enabled with search radii of 1 pixel and 1 channel, and we used a S/N threshold of 3 to determine whether the detected sources were reliable. We could achieve satisfactory masks with their native spatial and spectral resolutions for 12 galaxies, 17 galaxies required to rebin their spectra to 10\,$\rm km\, s^{-1}$,  
and 6 to rebin their spectra to 10\,$\rm km\, s^{-1}$ and in addition smooth spatially to 2\farcs5; for the remaining 6 galaxies, {\tt \texttt{SoFiA 2}} failed to find a coherent 3D mask.

Additionally, after a careful visual inspection of the masks for each cube, we manually increased the size of some masks with regions of interest and/or with regions with potential emission: galaxy \texttt{27516}, to cover the northern part of its disc, and galaxy \texttt{46746}, after visualising possible emission in its cube (but the flux is $\rm S/N<3$).
In the cases where we did not obtain a 3D mask with {\tt \texttt{SoFiA 2}}, we created an elliptical mask delimited by the size of the $d_{25}$, position angle (PA), and inclination (Incl.), and extending $\pm 150$ $\rm km\, s^{-1}$ around the optical velocity. The CO emission of these regions will be considered in the analysis as upper limits but may be relevant when compared to potential detections in SFR and stellar content.

Once we had masks for every galaxy, we calculated CO(1--0) fluxes integrating within their corresponding mask. After estimating their flux and respective uncertainty, we classified galaxies with $\rm S/N<2$ as ``non-detections'', galaxies with $\rm 2\leq S/N<5$ as tentative detections, and galaxies with $\rm S/N\geq 5$ as detections. 
We show in \cref{tab:ALMA_products} the results of this classification: 33 (80\%) are detected, 2 (5\%) are tentative, and 6 (15\%) of the galaxies are non-detected (with failed masking procedure).  

\subsubsection{Potential flux loss due to lack of short-spacings}

In Appendix \ref{appendix:fluxloss} we present the comparison of ALMA and IRAM 30m CO(1--0) \citep{Rodriguez2024} spectra and their derived integrated fluxes. We use a common channel width of 10\,\kms.
The shape of the profiles as well as central velocities agree, and integrated fluxes are usually within the uncertainties, especially at the high-flux end. The median ratio of the aperture corrected IRAM 30m flux with respect to the ALMA flux for tentative and detected galaxies is 1.02, with a median absolute deviation of 0.22, which confirms that in general no flux was lost as a result of missing short-spacings in the ALMA observations. However, in the low-flux regime, in a few cases the integrated fluxes disagree even if taking absolute flux scale uncertainties in both instruments. These differences are probably due to poor S/N and uncertain baseline subtraction in the single dish data, and not directly related to missing flux in the interferometric observations. The distribution in each channel of the ALMA datacube is usually below the imposed LAS of 15\arcsec.

\subsection{CAVITY and MaNGA IFU Data}
\label{sec:ifu}

In this section we present the IFU data used to obtain spatially resolved stellar, SFR and gas-phase metallicity maps, among other properties.
These data are from either PMAS/PPak observations  
 or compiled from MaNGA for objects within the ALMA CO-CAVITY sample. 
 
Observations for the CAVITY project were performed at the Calar Alto Observatory (CAHA) with the Postdam Multi Aperture Spectrograph (PMAS) in the PPak mode on the 3.5 meter telescope. The observation strategy, data reduction and data structure of the CAVITY PMAS/PPak data were presented in  \citet{Garcia_Benito2024}, and in the project overview \citet{Perez2024}. Data products were obtained using \texttt{pyPipe3D} \citep{Lacerda2022} for the CAVITY project, as introduced in \cite{Sanchez2024}. This is important because they follow the same prescriptions, such as consistent stellar population models and the same analysed emission lines.
The nominal Point Spread Function (PSF) of the observations is around 2\farcs5  \citep{Sanchez2016b}. 

A pixel size of 1\arcsec\ was chosen. 
In addition, 11 galaxies out the 41 galaxies in the ALMA CO-CAVITY sample were observed as part of Mapping Nearby Galaxies at APO \citep[MaNGA,][]{Bundy2015}, and they were not included for observations at CAHA. We compiled data products from MaNGA data release 17\footnote{\url{https://www.sdss4.org/dr17/MaNGA/MaNGA-data/MaNGA-pipe3d-value-added-catalog/}}, as 
\texttt{pyPipe3D}  was also used to obtain these data products \citep{Sanchez2022}. The PSF of the data is similar to that of PMAS/PPak, $\sim$ 2\farcs5, and pixel size is 0\farcs5 .

\referee{The PPak FoV is 74\arcsec $\times$ 64\arcsec \citep{Perez2024} and all targets are selected to be covered by at least their $d_{25}$, while for MaNGA the FoV is in the range 12 -- 32\arcsec\  \footnote{\url{https://www.sdss4.org/dr14/manga-instrument/}} and the target selection ensures a minimum of 1.5 $\times$ $R_{\rm eff}$ coverage. In our case the FoVs of the 11 galaxies with MaNGA data are 17\arcsec\ (1 object), 22\arcsec\ (2), 27\arcsec\ (2), and 32\arcsec\ (6) in diameter. Despite this difference in FoV, in all cases the regions of the galaxies that show significant emission are fully contained within their FoVs. Therefore, the differing spatial coverage does not affect our measurements or the conclusions of this work.}

\section{Results}
\label{sec:results}

\begin{figure*}
    \centering
        \sidecaption
    \includegraphics[width=\textwidth]{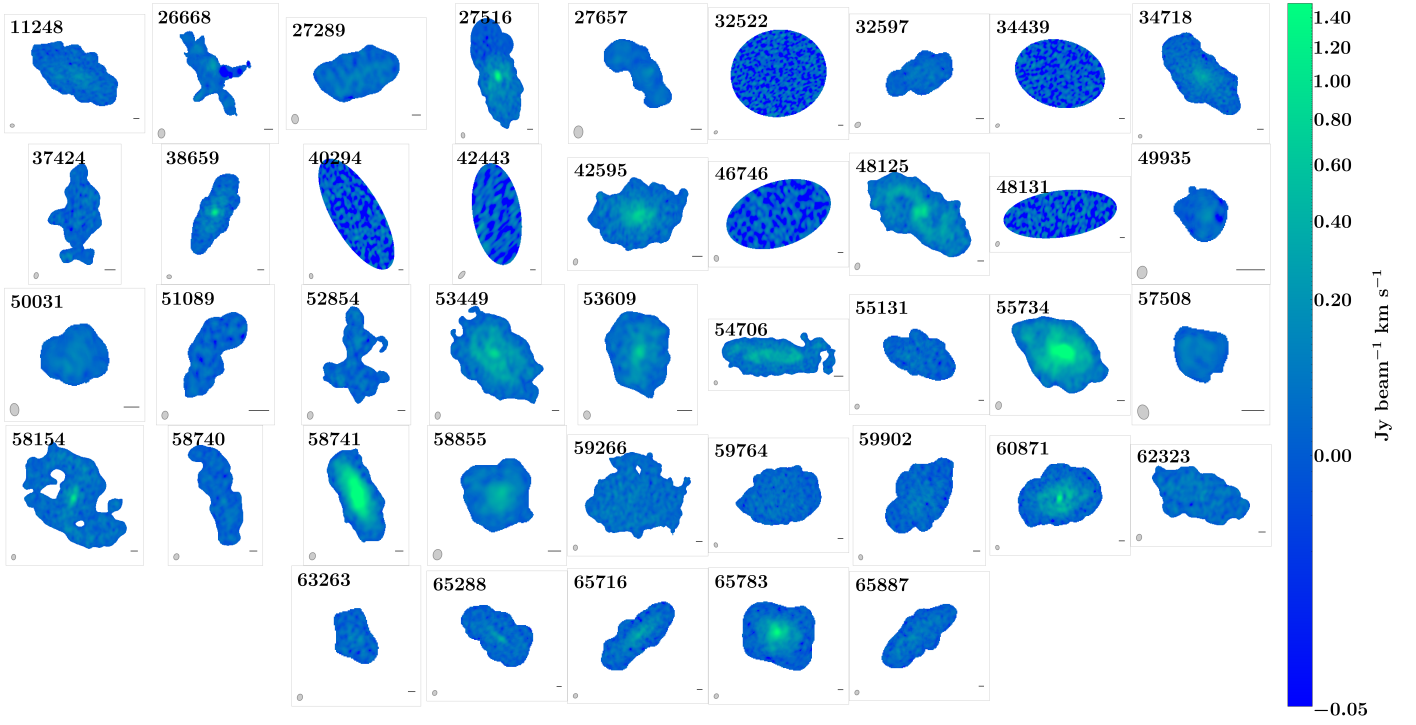}
    \caption[mugshot]{
    Moment-0 maps of the 41 ALMA CO-CAVITY VGs. In each panel, we show the CAVITY ID of the galaxy in the upper-left corner, \referee{the synthesized beam in the lower-left corner, }and a scale bar of 1\,kpc in the lower-right corner. The frame of each object corresponds to the rectangular field in \cref{fig:mugshot_all_galaxies}. \referee{The colour bar indicating the intensity range is shown on the right.}
    }
    \label{fig:mugshot_molecular_all_galaxies}
\end{figure*}

\referee{We present the molecular gas masses ($M_\mathrm{H_2}$) obtained from the ALMA CO-CAVITY data (\cref{subsec:molecularmass}), the main data products that will be publicly available as part of the ALMA CO-CAVITY Atlas at native angular resolution (\cref{subsec:atlas}), and the molecular surface densities ($\Sigma_{\rm H_2}$) obtained to a common 2\farcs5 angular resolution as $\Sigma_{\rm SFR}$ and $\Sigma_\star$ (\cref{subsec:resolved_physical_properties}). 
For a quick overview, \cref{fig:mugshot_molecular_all_galaxies} shows the CO(1--0) distribution (moment-0 maps) for all galaxies in the ALMA CO-CAVITY sample, following the outline and indicated FoVs in \cref{fig:mugshot_all_galaxies}. Finally, we present the main global scaling relations between $\Sigma_{\rm H_2}$ , $\Sigma_{\rm SFR}$ and $\Sigma_\star$: the Schmidt-Kennicutt law (SK), the molecular gas main sequence (MGMS), and the SFMS  (\cref{sec:scaling_relations}). }

\subsection{Molecular gas masses}
\label{subsec:molecularmass}

We compute $M_{\rm H_2}$ from the CO(1--0) luminosity $L^\prime_{\rm CO}$ using the relation from \citet{Solomon1997}:
$$L^\prime_{\rm CO} [{\rm K \, km\, s^{-1} pc^{2}}]= 3.25 \times 10^7\, S_{\rm CO} \nu_{\rm rest}^{-2} D_{\rm L }^{2} (1+z)^{-1}$$
where  $S_{\rm CO}$ is the CO(1--0) integrated flux in Jy\,\kms, 
$D_{\rm L }$ is the luminosity distance in Mpc, $z$ the redshift,
and $\nu_{\rm rest}$ is the rest frequency of the line in GHz.
We then calculate $M_{\rm H_2}$ in $\mathrm{M_\odot}$ as
$M_{\rm H_2}= \alpha_{\rm CO} L^\prime_{\rm CO}$, 
\noindent where $\alpha_{\rm CO}$ includes a factor 1.36 to account for the contribution of heavy elements. 

\cref{tab:unresolved} presents the molecular gas masses $M_{\rm H_2,\ ALMA}$ obtained from the ALMA data for the galaxies in ALMA CO-CAVITY sample. 
We present for comparison the masses obtained from our ALMA data ($M_{\rm H_2,ALMA}$) and those from the aperture-corrected IRAM 30m data ($M_{\rm H_2,\ 30m}$) in Appendix \ref{appendix:fluxloss} (see \cref{tab:molecular}).

\begin{figure}
    \centering
    \includegraphics[width=0.5\columnwidth]{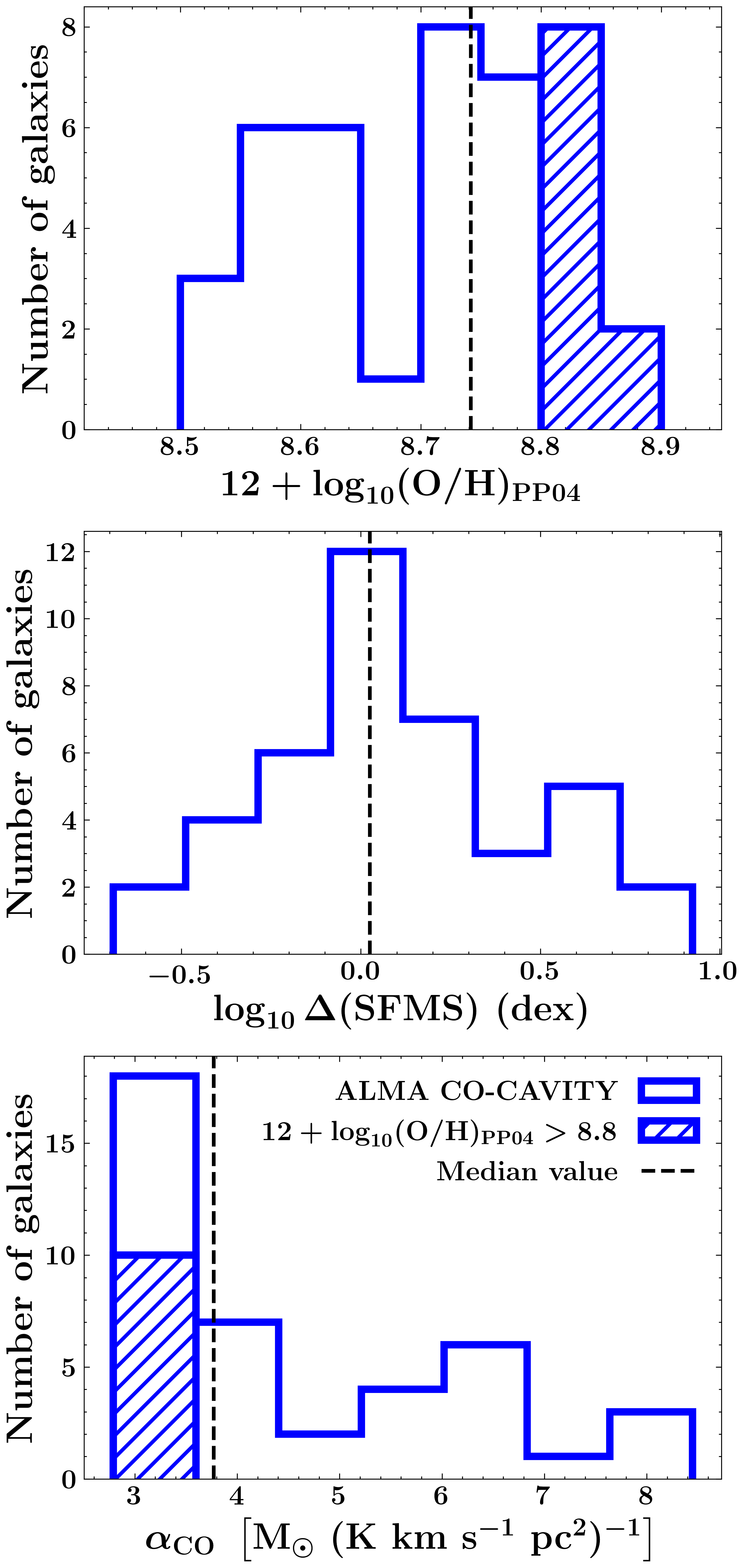} 
    \caption{
    Distribution of gas phase metallicity ($\rm 12+\log(O/H)$) (upper panel), distance to the SFMS ($\log\Delta({\rm SFMS})$) (mid panel), and the CO-to-H$_{2}$ conversion factor ($\alpha_{\rm CO}$) (bottom panel), for the ALMA CO-CAVITY sample. Hatched bins indicate the ten galaxies with $\rm 12+\log(O/H)>8.8$ (see \cref{subsec:molecularmass}).
    } 
    \label{fig:alpha_CO_DeltaSFMS_met}
\end{figure}

We use $\alpha_{\rm CO}$ for each galaxy as presented in \cref{tab:unresolved} to calculate $M_{\rm H_2}$. As in \citet{Rodriguez2024}, we use the \citet{Accurso2017} prescription which depends on the gas-phase metallicity and the distance to the main sequence, adopted from \citet{Janowiecki2020}, in the form:
$$    \log(\alpha_{\rm CO}) = 14.752 - 1.623[12 + \log({\rm O/H})] + 0.062 \log \Delta({\rm SFMS})
$$ 
\noindent where $\rm 12+\log(O/H)$ is the gas-phase metallicity from \cref{tab:galaxy_properties} 
and $\log \Delta({\rm SFMS})$ is the vertical distance to the SFMS.  
The uncertainty on the predicted log($\alpha_{\rm CO}$) is $\pm$0.165, being the metallicity the main driver of variations before $\Delta$(SFMS). 
For our sample, 12 + log (O/H) spans 0.4\,dex (from $8.5$ to $8.9$ dex) and log $\Delta$(SFMS) 1.61 dex (from $-0.68$ to $0.93$ dex). 
The distributions are shown in \cref{fig:alpha_CO_DeltaSFMS_met}. As stated by \citet{Accurso2017}, the applicability range in gas metallicity is $7.9 < \rm 12+\log(O/H) < 8.8$ (O3N2 indicator, \citealt{Pettini_Pagel2004}) and, above, it is suggested to calculate $\alpha_{\rm CO}$ using $\rm 12+\log(O/H)=8.8$.  Ten objects are over that limit (7 of them being just below 8.83) and we have set it to 8.8. 
We note that in all these cases the difference in $\alpha_{\rm CO}$ by fixing them to this limit is well within its uncertainty.
\citet{Accurso2017} also recommend the applicability range in $\Delta$(SFMS) of $-0.8 < \log \Delta({\rm SFMS}) < 1.3$, as is the case for all our galaxies.  In \cref{tab:unresolved} we show the $\Delta$(SFMS) obtained for each galaxy that have been considered, as well as the distance to the MGMS $\Delta$H$_2$MS, estimated following Eq. 7 of \cite{Janowiecki2020}.

\begin{table*}[]
    \centering
    \caption{Molecular gas masses of the ALMA CO-CAVITY sample}
    \begin{tabular}{ccccc}
    \hline\hline
       CAVITY ID  & log($M_{\rm H_2,\ ALMA}$) & $\alpha_{\rm CO}$ & $\log\Delta$H$_2$MS & $\log\Delta$SFMS  \\
                  & log([M$_\odot$])        &   ($\rm M_\odot\ (K\ km\ s^{-1}\ pc^2)^{-1}$)  & (dex) & (dex) \\
      (1) &         (2)  &                    (3)  &             (4)  &              (5) \\ \hline
    11248 & $8.966 \pm 0.017$ & 3.13 & 0.05 & 0.03 \\
    26668 & $8.43 \pm 0.10$ & 5.82 & -0.04 & -0.01 \\
    27289 & $8.68 \pm 0.03$ & 4.18 & -0.02 & -0.08 \\
    27516 & $9.166 \pm 0.011$ & 3.18 & -0.08 & -0.11 \\
    27657 & $8.59 \pm 0.05$ & 8.45 & 0.24 & 0.51 \\
        ...\\
    \hline
    \end{tabular}
    \tablefoot{
    (This table is truncated to the first 5 entries. The complete table is presented in its entirety in the online version of the publication.) (1) ID: CAVITY identifier in the CAVITY sample. (2) log($M_{\rm H_2,\ ALMA}$): the molecular gas mass from ALMA data in decimal log scale. 
    (3) $\alpha_{\rm CO}$: The CO to H$_2$ conversion factor obtained as in \cite{Rodriguez2024}, see also \cref{sec:sample}. (4) $\log\Delta$H$_2$MS: the distance to the MGMS of \citet{Janowiecki2020}. (5) $\log\Delta$SFMS: the distance to the SFMS main sequence of \citet{Janowiecki2020}.
    }
    \label{tab:unresolved}
\end{table*}

\begin{figure}[]
    \centering
        \includegraphics[width=0.6\columnwidth]{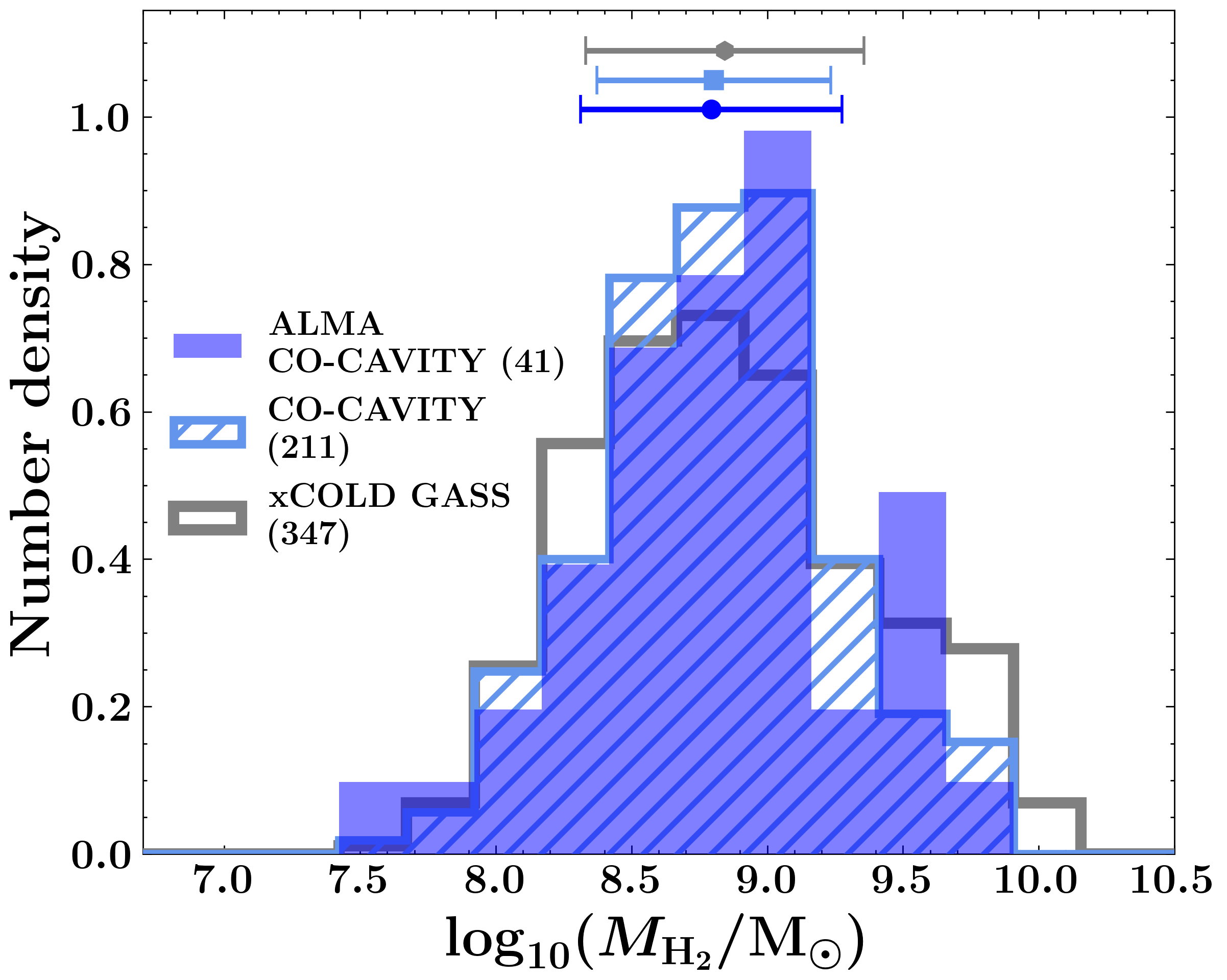}
        \includegraphics[width=0.6\columnwidth]{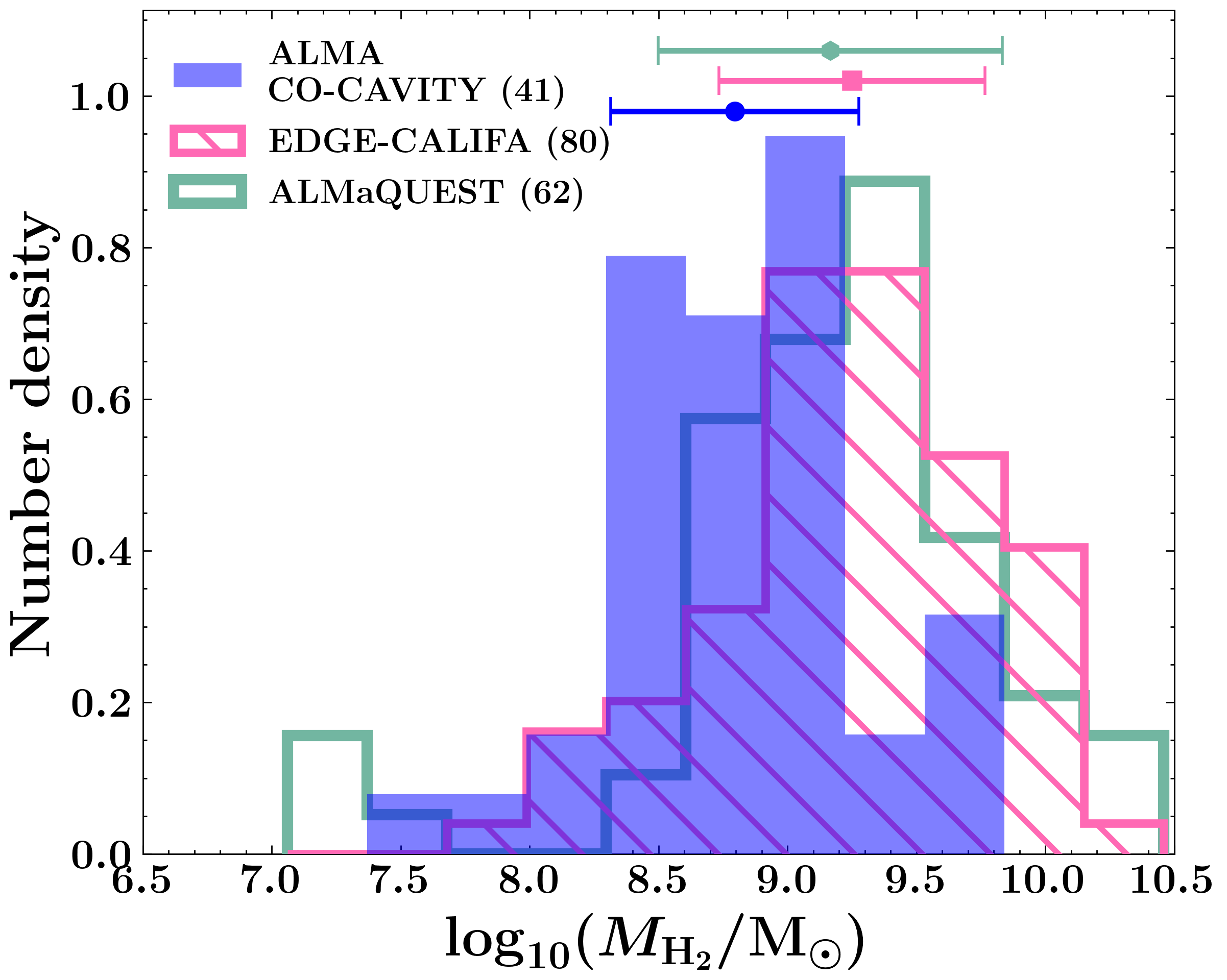}
    \caption{
    $M_{\rm H_2}$ distribution of the ALMA CO-CAVITY sample (blue filled histogram), compared to: (Top) unresolved studies CO-CAVITY (blue hatched), xCOLD GASS samples (dark grey unfilled); (Bottom) resolved studies EDGE-CALIFA (magenta hatched), and ALMaQUEST (green unfilled) samples. Above each histogram we show bars with the mean values and standard deviations.}
    \label{fig:MH2_histograms}
\end{figure}

In \cref{fig:MH2_histograms} we present the $M_{\rm H_2}$ distribution for the ALMA CO-CAVITY, CO-CAVITY, xCOLD GASS, EDGE-CALIFA, and ALMaQUEST samples. The means and standard deviations are very similar: $8.8\pm0.5$, $8.8\pm0.4$ and $8.8\pm0.5$ dex for the ALMA CO-CAVITY, CO-CAVITY and xCOLD GASS samples, respectively; and with K-S test p-values of 0.97 and 0.71 when comparing the ALMA CO-CAVITY distribution with the CO-CAVITY and xCOLD GASS distributions, respectively.  
This was expected between ALMA CO-CAVITY and CO-CAVITY samples, since the galaxies in the former are mostly extracted from the latter, but also the same results from the comparison with the xCOLD GASS survey. The mean value of $\rm \log\Delta H_2MS$ is $0.0\pm0.3$ dex for the ALMA CO-CAVITY sample, similarly to what is found for the void and control samples in \citet{Rodriguez2024}, whose mean values are $0.0\pm0.4$ and $-0.2\pm0.4$\,dex, respectively.
However, we observe generally smaller log($M_{\rm H_2}$) by about 0.5\,dex in the ALMA CO-CAVITY distribution compared to those of the EDGE-CALIFA and ALMaQUEST samples. The mean values are $9.2\pm0.5$ and $9.2\pm0.7$ dex for the EDGE-CALIFA and ALMaQUEST samples, respectively; and the p-values of K-S tests comparing both distributions with that of ALMA CO-CAVITY are $1.54\times 10^{-5}$ and $1.44\times 10^{-4}$, respectively for EDGE-CALIFA and ALMaQUEST comparisons. This offset can be understood from the fact that  
VGs are less massive  
and the MGMS.  
We note that using a fixed value of $\alpha_{\rm CO}=4.3\ \mathrm{M_{\odot}\ (K\ km\ s^{-1}\ pc^{2})^{-1}}$ as in EDGE-CALIFA and ALMaQUEST does not substantially change the offset. 

\subsection{ALMA CO-CAVITY Atlas}
\label{subsec:atlas}

\begin{figure*}
    \centering
    \sidecaption
    \includegraphics[width=0.7\linewidth]{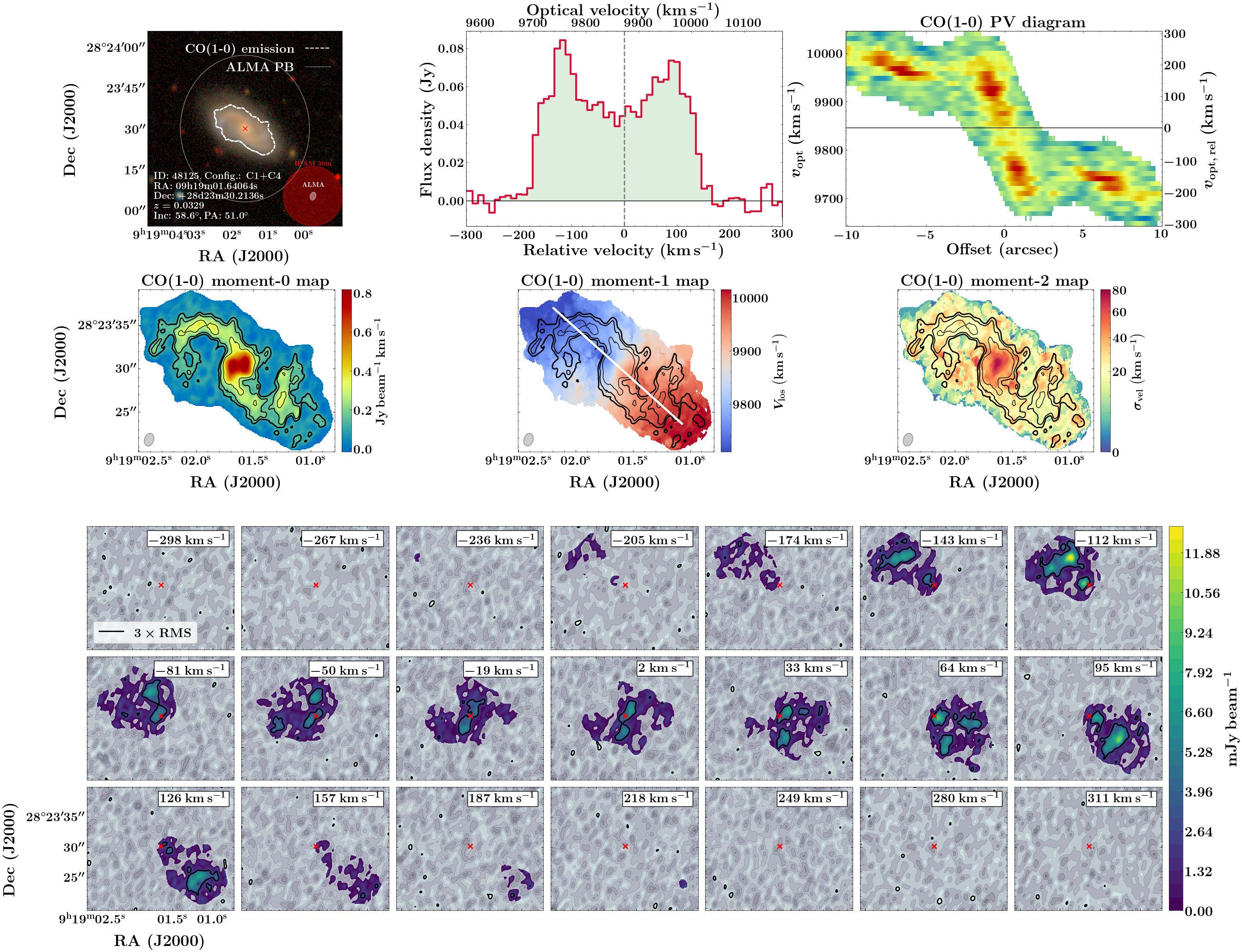} 
    \caption{
    Example of the ALMA CO-CAVITY Atlas data products at native angular resolution, for galaxy CAVITY ID \texttt{48125}: Upper row, from left to right: Optical image (DECaLS) with the galaxy information, configuration used, ellipses with ALMA and IRAM 30m beam sizes (HPBW), the centred circle indicating the ALMA primary beam, the dashed contour the CO(1--0) mask, and the red cross the optical photometric centre of the galaxy; CO(1--0) ALMA spectrum with the green area showing the range of velocities with detected emission, and the vertical dashed line showing the expected position of the CO(1--0) line based on its redshift from SDSS ($cz$); Position-velocity (PV) diagram along the semimajor axis of the galaxy as indicated in the moment-1 map with a white line. Mid row, from left to right: Moment-0, 1 and 2 maps of the CO(1--0) emission, with moment-0 contours at S/N = 2, 3 and 5 in all of them; the moment-1 and 2 maps were obtained applying a 1$\sigma$ clipping; Bottom row: Channel map indicating the masked datacube, $\pm 300\ {\rm km\ s^{-1}}$ around the centre of the emission, with black contours showing the 2, 3 and 5$\times {\rm RMS_{chan}}$ level, where $\rm RMS_{chan}$ is that presented in \cref{tab:ALMA_products}, and the red cross showing the optical photometric centre of the galaxy.}
    \label{fig:moms}
\end{figure*}

We obtained maps for the molecular gas morphology, kinematics and velocity dispersion in this sample of galaxies, with resolutions of $\sim 1\arcsec$ (1--2\,kpc). 
In \cref{fig:moms} we present an example of the data products we obtained at native angular resolution that compose the ALMA CO-CAVITY Atlas, in particular for galaxy CAVITY ID \texttt{48125}. 
This includes the CO(1--0) spectrum, moment-0, 1 and 2 maps of the CO(1--0) emission, channel maps, as well as the position-velocity (PV) diagram along the PA of the galaxies. For each galaxy we also include an optical image of the galaxy, created using 3 photometric bands from DECaLS ($g$, $r$, and $i$ or $z$). It contains the extent of the ALMA observations' primary beam (HPBW 54\arcsec), the synthesized beam of the ALMA data and the beam of the IRAM 30m telescope (HPBW 22\arcsec) (if observations exist, see Appendix \ref{appendix:fluxloss}), the optical photometric centre of the galaxy, the CO(1--0) mask and other information such as the CAVITY ID, the ALMA configuration(s) used, the galaxy coordinates, redshift, inclination and PA.  
The moment maps were obtained using the task {\tt immoments}, using the channel range where the emission line was detected, and using the obtained mask regions. 
The PV plot was done using the task {\tt impv} with the same mask. 
The ALMA CO-CAVITY Atlas with the information for all galaxies in the sample as \cref{fig:moms} is publicly available in electronic form.

\subsection{Resolved maps at common resolution of \texorpdfstring{2\farcs5}{2.5''}}
\label{subsec:resolved_physical_properties}

\begin{figure*}
    \centering
    \sidecaption
    \includegraphics[width=0.7\textwidth]{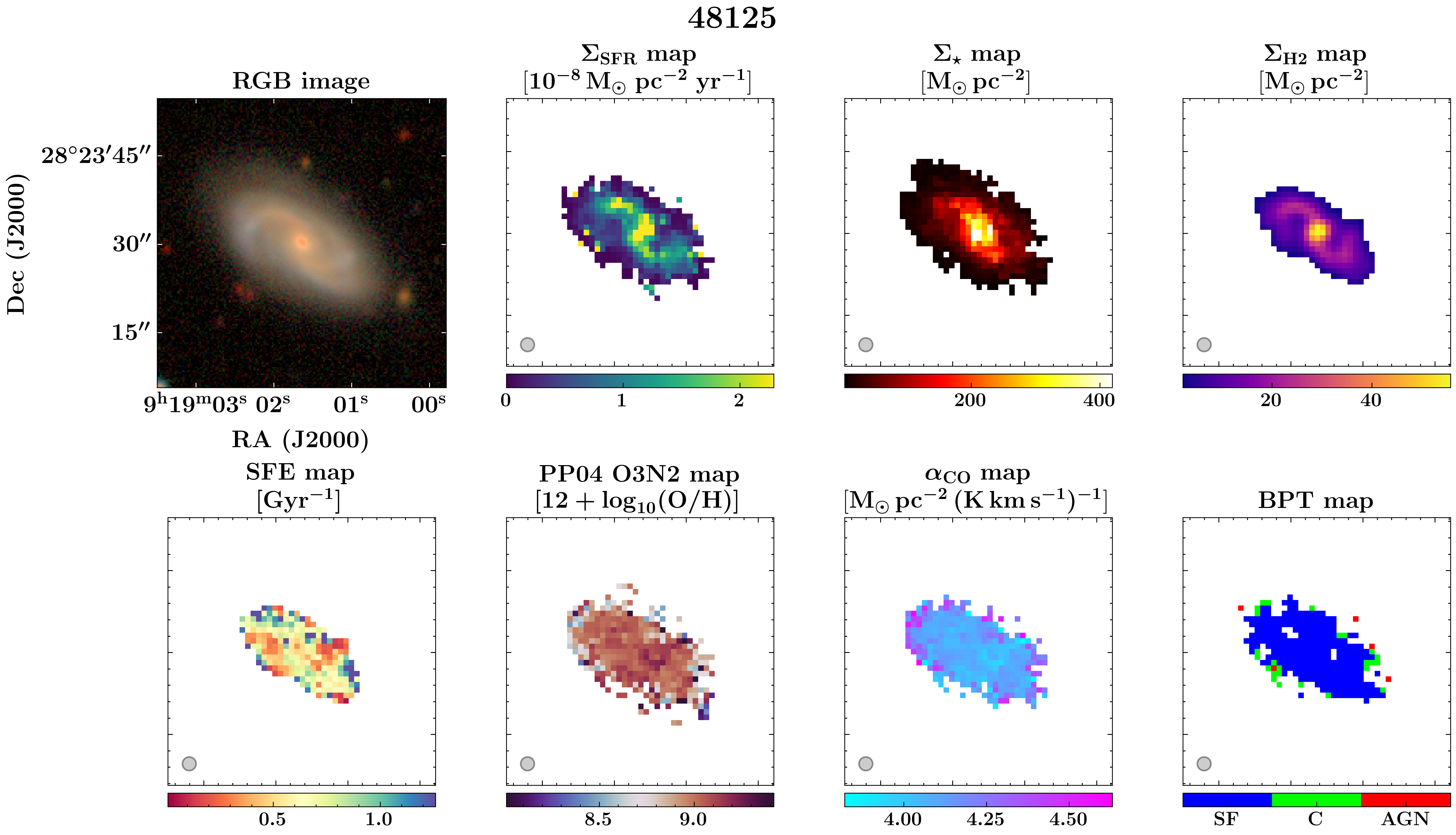}
    \caption{
    Example of the ALMA CO-CAVITY resolved data products at 2\farcs5 angular resolution, for galaxy CAVITY ID \texttt{48125}. In the upper row (from left to right): optical image (as in \cref{fig:moms} but with the same field of view as the other maps), SFR surface density ($\Sigma_{\rm SFR}$) map, stellar surface density ($\Sigma_\star$) map, molecular gas surface density ($\Sigma_{\rm H_2}$) map. In the lower row: star formation efficiency (SFE) map, PP04 03N2 gas-phase abundances ($\rm 12+\log(O/H)$) map, conversion factor ($\alpha_{\rm CO}$) map, and BPT map (SF for star-forming, C for composite and AGN  
    for AGN-like region).
    Colour scales are shown at the bottom of each map. This figure can be found in electronic form for all galaxies in the ALMA CO-CAVITY sample.}
    \label{fig:allmaps}
\end{figure*}

\cref{fig:allmaps} shows an example of the ALMA CO-CAVITY resolved data products at a common 2\farcs5 angular resolution of the IFU data, for galaxy CAVITY ID \texttt{48125}. \cref{fig:allmaps} includes the optical image, SFR surface density ($\Sigma_{\rm SFR}$) map, stellar surface density ($\Sigma_\star$) map, molecular surface density ($\Sigma_{\rm H_2}$) map, SFE map, gas-phase metallicity ($\rm 12+\log(O/H)$) map, conversion factor ($\alpha_{\rm CO}$) map, and the BPT diagram map. 
The $\Sigma_\star$ maps were obtained as a direct product of \texttt{pyPipe3D}, fitting single stellar populations with the \texttt{MaStars\_sLOG} stellar library. The $\Sigma_{\rm SFR}$ was estimated pixel-by-pixel using the $\Halpha$ flux corrected for extinction provided by \texttt{pyPipe3D}, and following the estimator from \cite{KennicuttEvans2012}, adapted to the Chabrier IMF. Similarly, the gas-phase abundances were estimated pixel-by-pixel using \cite{Pettini_Pagel2004} [\ion{N}{II}]/H$\alpha$ and [\ion{O}{III}]/H$\beta$ line ratios (O3N2) indicator.  Finally, the $\Sigma_{\rm H_2}$ maps were obtained pixel-by-pixel following the same procedure described in \cref{subsec:molecularmass}, but to a common pixel scale and resolution of the IFU maps. 
More details including the corrections applied to the maps, the estimation of upper limits and the homogenisation of the data will be provided in Paper II.

\subsection{Global scaling relations: SK, MGMS, SFMS}
\label{sec:scaling_relations}

\begin{figure*}[h]
    \centering
    \includegraphics[width=0.9\textwidth]{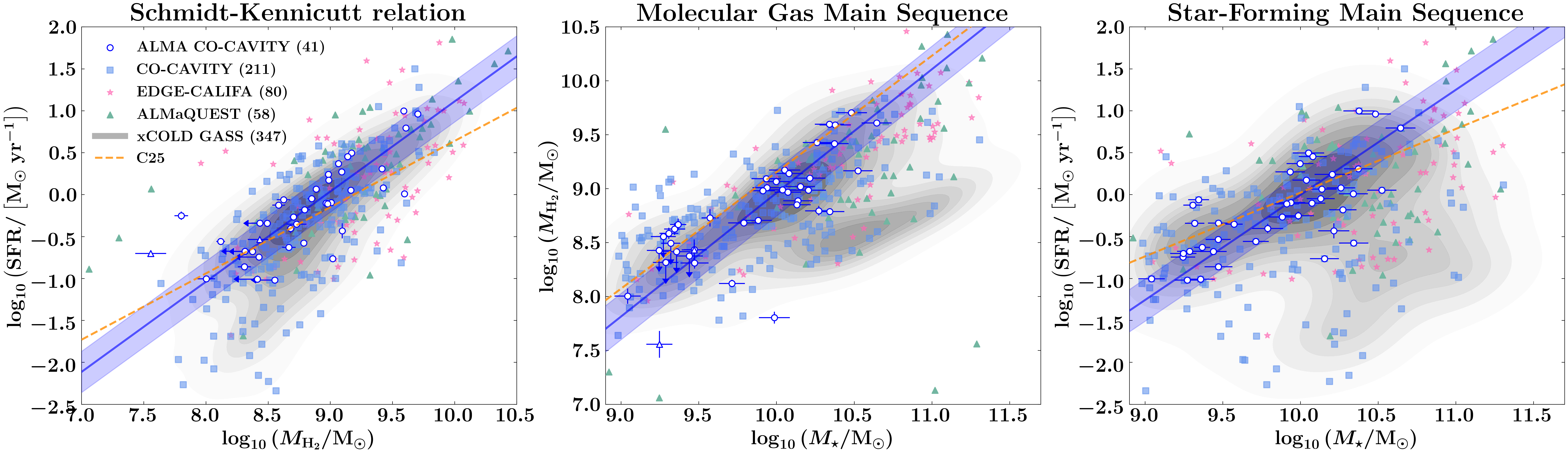}
    \caption{
    Global scaling relations of the ALMA CO-CAVITY sample: (\textit{left}) Schmidt-Kennicutt relation, (\textit{middle}) molecular gas main sequence, (\textit{right}) star-forming main sequence. Data points with error bars are presented for all galaxies in the ALMA CO-CAVITY sample, and arrows show upper limits. Unfilled triangles are tentative detections in the ALMA CO-CAVITY sample. The orthogonal distance regression (ODR) fits to the ALMA CO-CAVITY sample are shown as solid lines and the bands show the obtained 1$\sigma$ scatter of the fit. The fit parameters are indicated in \cref{tab:scaling_relations}. Data points are shown for comparison for CO-CAVITY (filled blue squares), EDGE-CALIFA (pink stars) and ALMaQUEST (filled green triangles), and the distribution for xCOLD GASS. We also added the fits (orange dashed lines) from iEDGE \citep{2025A&A...699A.367C} (converted to Chabrier IMF) for comparison. 
    }
    \label{fig:three_relations_unresolved}
\end{figure*}

In \cref{fig:three_relations_unresolved} we present the global Schmidt-Kennicutt (SK, \citealt{1998ARA&A..36..189K}), Molecular Gas Main Sequence (MGMS, \citealt{2016MNRAS.462.1749S}) and Star-forming Main Sequence (SFMS, \citealt{Brinchmann2004}) scaling relations for the galaxies of ALMA CO-CAVITY, in comparison with previous surveys including global data: CO-CAVITY \citep{Rodriguez2024}, xCOLD GASS \citep{Saintonge2017}, and iEDGE \citep{2025A&A...699A.367C}.
 We also added global data points for the resolved studies ALMaQUEST and EDGE-CALIFA, to show how their distributions compare to ours. 
To make a fair comparison among the different surveys, we make sure that
the radius of integration cover the whole galaxy, 
we convert to the same IMF (Chabrier), 
and that we all account for the factor 1.36 due to helium and heavier elements.
We use individual $\alpha_{\rm CO}$ for each galaxy as in CO-CAVITY and xCOLD GASS, but we note that EDGE-CALIFA and ALMaQUEST used a fixed Galactic $\alpha_{\rm CO}$. Again, considering a fixed $\alpha_{\rm CO}$ in our case does not change the results.

 \begin{table}[]
\caption{Best-fit relations and intrinsic scatter for the SK, MGMS, and SFMS relations.}
\label{tab:scaling_relations}

    \begin{tabular}{llccc}
    \hline \hline  
    Relation & Sample & Slope & Intercept & Scatter \\
             &      &       &           &  (dex)  \\
    \hline
    SK  & ACC VGs &$1.08 \pm 0.13$ & $-9.6 \pm 1.1$ & 0.25 \\ 
        &  CC VGs (SF) &$0.84 \pm 0.06$ & $-7.6 \pm 0.5$ & 0.17 \\ 
        &  CC CS (SF) &$0.72 \pm 0.04$ & $ -6.5 \pm 0.4$ & 0.13 \\ 
        & iEDGE (SF)& $0.79 \pm 0.02$ & $ -7.3 \pm  0.2 $ & $0.13$ \\ \hline

    MGMS &  ACC VGs&$1.15 \pm 0.13$ & $-2.5 \pm 1.3$ & 0.21 \\ 
         &  CC VGs (SF) &$0.89 \pm 0.06$ & $0.1 \pm 0.6$ & 0.19 \\ 
         &  CC CS (SF) &$0.88 \pm 0.04$ & $0.1 \pm 0.4$ & 0.13 \\ 
         &  iEDGE (SF)& $1.08 \pm 0.03$ & $ -1.7 \pm 0.3$ & $0.15$ \\ \hline

    SFMS &  ACC VGs&$1.25 \pm 0.16$ & $-12.5 \pm 1.6$ & 0.25 \\
         &  CC VGs (SF) &$0.77 \pm 0.04$ & $-7.8 \pm 0.4$ & (0.13) \\ 
         &  CC CS (SF) &$0.63 \pm 0.03$ & $-6.4 \pm 0.3$ & (0.13) \\ 
         &  iEDGE (SF) & $0.76 \pm 0.02$ & $ -7.6 \pm 0.2$ & $0.06$  \\

    \hline
    \end{tabular}
    \tablefoot{
Samples are: 1) ACC VGs: ALMA CO CAVITY galaxies, 2) CC VGs (SF): The subsample  of CO-CAVITY VGs on the SFMS \citep[|sSFR-SFMS| $<$ 0.3 dex][]{Rodriguez2024}. The scatter is shown in parenthesis in the SFMS, as it is biased by the selection of the star-forming subsample. 3) CC CS (SF): Same as 2) but for the CO-CAVITY control sample. 4) iEDGE: `star forming' group of the iEDGE sample following  \citet{2025A&A...699A.367C}. The fits are obtained with ODR, except for \citet{2025A&A...699A.367C}. It has been converted to the Chabrier IMF. 
}
\end{table}

The orthogonal distance regression (ODR) fits for the ALMA CO-CAVITY sample are presented in \cref{tab:scaling_relations}. 
From these relations and the data points in \cref{fig:three_relations_unresolved} we can see that ALMA CO-CAVITY galaxies are, in general, located in the active star-forming part of these plots and with a relatively normal amount of molecular gas. 
\referee{We find the lowest (orthogonal) scatter in the  MGMS, 0.21, followed by SK and SFMS, from 0.21 to 0.25\,dex, respectively. The SFMS has the largest uncertainties in the fit parameters. }
We compare the fits with those of the subsample  of CO-CAVITY VGs on the SFMS  \citep[|sSFR-SFMS| $<$ 0.3 dex,][]{Rodriguez2024}, since ALMA CO-CAVITY is dominated by this kind of galaxies, and we find that there is good agreement taking into account the applicability range and their scatter.

A similar finding is reached for galaxies in denser environments. First, we use the subsample of CO-CAVITY control sample (based on xCOLD GASS, but removing VGs) on the SFMS with the same criterion \citep{Rodriguez2024}. 
Second, we compare with the recent iEDGE sample \citep{2025A&A...699A.367C} which contains EDGE-CALIFA data plus an extension including single-dish APEX and ACA data, totaling data for over 600 galaxies in the local Universe. Again, we compare with the 'star forming' group as classified by \citet{2025A&A...699A.367C}. In both cases the fits (see \cref{fig:three_relations_unresolved} and \cref{tab:scaling_relations}) are compatible with ours, being the closest match that of the MGMS.

\section{Discussion}
\label{sec:discussion}

\referee{We find that VGs contain significant amounts of molecular gas, in agreement with previous studies (e.g. \citealt{Dominguez_Gomez2022}; \citealt{Rodriguez2024}). For the first time, we provide spatially resolved information on the molecular gas morphology and kinematics for a relatively large sample of VGs. The molecular gas exhibits a range of morphologies: while the majority of galaxies show centrally concentrated distributions, several display structures such as spiral arms, ring features, and bars. In most cases, the molecular gas distribution correlates well with both $\Sigma_\star$ and $\Sigma_{\mathrm{SFR}}$. In this sense, qualitatively, we do not find major differences with surveys of galaxies in other environments. However, a more quantitative comparison beyond the scope of this work is required to assess potential differences. In particular, analyses of radial distributions \citep[][]{Bolatto2017,2024ApJ...964..120P} and the derivation of rotation curves \citep[e.g.][]{2022ApJ...934..173S} would provide further insight. We can also distinguish galaxies with asymmetric molecular gas distributions, low surface brightness, or even non-detections, typically corresponding to lower-mass systems, likely reflecting lower metallicities. A detailed quantitative characterization of these properties will be presented in future work.}

\referee{Establishing a robust baseline for comparison with other samples is critical. Such comparisons require consistent assumptions, including the use of the same prescriptions for key parameters (e.g. IMF, $\alpha_{\mathrm{CO}}$), similar spatial resolution and pixel sizes, and homogeneous data processing pipelines. It is equally important to distinguish intrinsic differences between samples, such as $M_\star$, size, star-forming properties, large-scale structure, degree of isolation, among others. We have imaged a sample of galaxies representative of star-forming systems in the least dense regions of the Universe, which we have seen that are predominantly composed of isolated objects, especially when compared to other samples. As expected for a VG sample, it is characterised, on average, by lower stellar masses ($M_\star$) and lower molecular gas masses ($M_{\mathrm{H_2}}$) compared to galaxies in denser environments.}

\referee{
{\it Dependence of global scaling relations on LSE}. 
We derive global scaling relations and find that, when restricted to star-forming galaxies, VGs follow relations consistent with those observed in other environments. Within the uncertainties, these relations appear largely insensitive to the LSE, at least to first order. This is consistent with previous results showing no significant environmental dependence of the SFE \citep{Rodriguez2024}, and with studies such as \citet{Ricciardelli2014}, which find that the SFMS remains remarkably invariant across environments when only star-forming galaxies are considered. 
It is important to note that we find the lowest scatter in the  MGMS, followed by SK and SFMS. This is in agreement with SFMS being the least physically fundamental and probably resulting as the by-product of the MGMS and SK relations \citep[e.g.][]{2019ApJ...884L..33L}.
Our sample, composed mostly of relatively isolated galaxies in the lowest density environments, provides a useful baseline for star-forming systems. This baseline can be used for comparison with other populations (from isolated, \citealt{2011A&A...534A.102L} to interacting systems and mergers, \citealt{2018ApJ...866...77E}). The scaling relations presented in \cref{tab:scaling_relations} may also be used to estimate  $M_{\mathrm{H_2}}$ to quantify enhancements or deficiencies, as it is typically more observationally expensive to obtain than $M_\star$ (and SFR).
A key next step is to determine whether these scaling relations hold at resolved (kpc) scales, which will be addressed in the forthcoming Paper II.}

\referee{
{\it Relation between SFHs and molecular gas reservoirs. } 
It has been found that VGs experience in average slower SFHs, potentially linked to gradual gas assembly \citep{Dominguez_Gomez2023}. Here we address the question of whether a slower SFH implies a larger molecular gas reservoir at present, as suggested by \citet{2026A&A...706A.265S} for the the VG CG-910. \citet{2026A&A...706A.265S} obtained resolved interferometric CO observations of this VG with angular resolutions of $\sim$ 3\arcsec (or 3\,kpc). They found an asymmetric molecular gas distribution and a significantly large molecular gas depletion timescale ($t_{\mathrm{dep}}$ $\sim$ 36\,Gyr), that they argued to be indicative of the slow evolutionary processes typical of void environments. Its relatively high stellar, molecular masses and SFR, (log($M_\star$) = 10.33, log($M_{\mathrm{H_2}}$) = 10.08, and SFR = 0.33\,M$_\odot$ \,yr$^{-1}$, would place it 1\,dex  above the MGMS but on the SFMS, and nearly outside the distribution of data-points of both the ALMA CO-CAVITY and CO-CAVITY samples, implying an abnormally large molecular gas content. These results support the picture that, although it is possible that some VGs can be molecular gas-rich yet inefficient at converting molecular gas into stars, this behavior is not typical and rather represents an exception in the void population.}

\section{Summary and Conclusions}
\label{sec:conclusions}

In this work we have presented the ALMA CO-CAVITY project, the first interferometric CO(1--0) survey targeting a representative sample of 41 void galaxies (VGs). This dataset offers an unprecedented, resolved view of molecular gas and its relation to SF and stellar content in the most isolated, low-density environments of the nearby Universe. We provide an homogeneous set of data products, including fully calibrated data cubes, moment maps, and PV diagrams, enabling robust pixel-to-pixel analyses of the physical processes governing galaxy evolution in extreme large-scale underdensities. By combining ALMA observations at typical resolutions of $\sim$1\,kpc with optical IFU data from CAVITY —either from PMAS/PPak or compiled from MaNGA— we have assembled spatially resolved maps of molecular gas, stellar properties, SF, and gas-phase metallicity at a common angular resolution. 

The ALMA CO-CAVITY sample is representative of the VG population while maintaining continuity with global galaxy surveys such as xCOLD GASS and iEDGE. In contrast, existing resolved CO surveys (e.g. ALMaQUEST and EDGE-CALIFA) primarily probe more massive systems in denser environments, highlighting the importance of ALMA CO-CAVITY in expanding the parameter space toward lower masses, higher isolation, and minimal environmental influence. 
Since VGs are predominantly isolated and dynamically undisturbed,
  they offer a clean baseline in the
  absence of external triggers. The global scaling relations established here, and resolved ones in future studies, therefore serve as baseline references with reduced environmental contamination, providing a benchmark for interpreting galaxy properties across the full range of environments.

From the global relations, i.e. the SFMS, SK and MGMS, the obtained fits are characterised with slopes close to 1 or slightly superlinear. 
We find that the (orthogonal) scatter for MGMS is the smallest, 0.21\,dex, followed by the SK and SFMS.
We obtain the relations for the star-forming VGs of (IRAM 30m) CO-CAVITY and the fits are compatible with those of ALMA CO-CAVITY. 
The fits for VGs (both ALMA CO-CAVITY and CO-CAVITY) are also consistent with samples of galaxies that are in general in other large-scale environments (CO-CAVITY control sample, \citealt{Rodriguez2024}; iEDGE sample, \citealt{2025A&A...699A.367C}), when we restrict them to star-forming galaxies within the SFMS. Although from global relations we cannot distinguish any dependence of the relations with the environment, we need to further probe this using the higher resolution of our data. 

This introductory paper lays the foundation of the project for a series of forthcoming studies exploiting this unique dataset. Future work will investigate in detail the resolved scaling relations that connect molecular gas, SF, and stellar structure; the role of local and large-scale environments in shaping galaxy properties; the radial distributions of gas and star-forming activity; and the mechanisms driving star-formation regulation and quenching in VGs. Together, these studies will build a comprehensive picture of how galaxies grow and evolve in the most pristine environments of the cosmic web.

\begin{acknowledgements}
 \referee{We thank the anonymous referee for carefully reviewing our manuscript. We truly appreciate the assessment of our work and the constructive comments provided, which have helped us to improve the paper.} 
This paper makes use of the following ALMA data: ADS/JAO.ALMA\#2022.1.00482.S. ALMA is a partnership of ESO (representing its member states), NSF (USA) and NINS (Japan),
together with NRC (Canada), MOST and ASIAA (Taiwan), and KASI (Republic of Korea), in
cooperation with the Republic of Chile. The Joint ALMA Observatory is operated by
ESO, AUI/NRAO and NAOJ.
The project leading to this publication has received support from ORP, that is funded by the European Union's Horizon 2020 research and innovation programme under grant agreement No 101004719.
Based on observations collected at the Centro
Astronómico Hispano en Andalucía (CAHA) at Calar Alto, operated jointly by Junta de Andalucía and Consejo Superior de Investigaciones Científicas (IAA-CSIC). 
This work is based on observations carried out with the IRAM 30\,m telescope. IRAM is supported by INSU/CNRS (France), MPG (Germany) and IGN (Spain).
This project makes use of the MaNGA-Pipe3D dataproducts. We thank the IA-UNAM MaNGA team for creating this catalogue, and the Conacyt Project CB-285080 for supporting them.
We acknowledge financial support by the research projects AYA2017-84897-P, PID2020-113689GB-I00, PID2020-114414GB-I00, PID2023-149578NB-I00, PID2023-150178NB-I00 financed by MCIN/AEI/10.13039/501100011033, the project A-FQM-510-UGR20 financed from FEDER/Junta de Andaluc\'{\i}a-Consejer\'{\i}a de Transformaci\'on Econ\'omica, Industria, Conocimiento y Universidades/Proyecto and by the grants P20-00334 and FQM108, financed by the Junta de Andaluc\'{\i}a (Spain). 
DE acknowledges support from a Beatriz Galindo senior fellowship (BG20/00224) from the Spanish Ministry of Science and Innovation.
SBD acknowledges financial support from the grant AST22.4.4, funded by Consejería de Universidad, Investigación e Innovación and Gobierno de España and Unión Europea –- NextGenerationEU, and project PID2021-122544NB-C43.
RGB acknowledges financial support from the Severo Ochoa grant CEX2021-001131-S funded by MCIN AEI/10.13039/501100011033 and PID2022-141755NB-I00.
TRL acknowledges support from Ram\'on y Cajal fellowship (RYC2023-043063-I, financed by MCIU/AEI/10.13039/501100011033 and by the FSE+).
GTR acknowledges financial support from the research project PRE2021-098736, funded by MCIN/AEI/10.13039/501100011033 and FSE+.
YGK acknowledges financial support from PREP2023-001684 funded by MCIU/AEI/10.13039/501100011033 and the FSE+. 
MAF and PVB acknowledge support from the Emergia program (EMERGIA20\_38888) from Consejería de Universidad, Investigación e Innovación de la Junta de Andalucía.

This work made use of the following software packages: \texttt{astropy} \citep{astropy:2022}, \texttt{Jupyter} \citep{2007CSE.....9c..21P, kluyver2016jupyter}, \texttt{matplotlib} \citep{Hunter:2007}, \texttt{numpy} \citep{numpy}, \texttt{pandas} \citep{mckinney-proc-scipy-2010, pandas_10697587}, \texttt{python} \citep{python}, \texttt{scipy} \citep{2020SciPy-NMeth, scipy_10909890}, \texttt{astroquery} \citep{2019AJ....157...98G, astroquery_10799414}, \texttt{scikit-learn} \citep{scikit-learn, sklearn_api, scikit-learn_10666857}, and CARTA (Cube Analysis and Rendering Tool for Astronomy) software (DOI \url{https://zenodo.org/records/15172686} – \url{https://cartavis.org}). 
This research has made use of the NASA/IPAC Extragalactic Database, which is funded by the National Aeronautics and Space Administration and operated by the California Institute of Technology. Funding for SDSS-III has been provided by the Alfred P. Sloan Foundation, the Participating Institutions, the National Science Foundation, and the U.S. Department of Energy Office of Science. The SDSS-III Web site is \url{http://www.sdss3.org/}. The SDSS-IV site is \url{http://www.sdss.org}. This project used data obtained with the Dark Energy Camera (DECam), which was constructed by the Dark Energy Survey (DES) collaboration.

\end{acknowledgements}

\bibliographystyle{aa.bst}
\bibliography{references.bib}

\onecolumn

\appendix
\section{Information of ALMA CO-CAVITY observations and properties of ALMA data cubes}
\label{appendix:informationObservations}

 \cref{tab:ALMA_observations} summarises the observational and processing characteristics of the ALMA datasets used in this work. We list the Member OUS (Observing Unit Set) identifiers, which specify the individual data units within the ALMA project, as well as the corresponding Scheduling Block (SB) names. We also present the ALMA array configuration employed for each observation. Then we provide the groups IDs which associates galaxies observed within the same SBs, and share flux and bandpass calibration. We report in this table the synthesised beam sizes of the data products as obtained by the CASA pipeline version used to calibrate and image each dataset by the observatory. For clarity, we separate the datasets obtained in the extended (i.e. C$-3$ and C$-4$) configurations from those observed in the more compact C$-1$ configuration.

 \cref{tab:ALMA_products} lists the main properties of the final ALMA data cubes used in this work for each galaxy in the ALMA CO-CAVITY sample, after data combination has been performed, specifying the ALMA array configuration(s) used to generate the final data cubes. We report the RMS noise measured in the cubes for a common channel width of 10 km s$^{-1}$. We also provide the final synthesised beam parameters for these combined data cubes. We provide a detection label assigned to each source as explained in \cref{sec:alma}. A source is considered detected when the signal-to-noise ratio (S/N) of the CO(1--0) integrated flux within the masked region exceeds 5; cases with $3 < \mathrm{S/N} < 5$ are classified as tentative detections, and those below this threshold are listed as non-detections.

\begin{table*}[h!]
    \centering
    \caption{Information of ALMA CO-CAVITY observations.}
    \label{tab:ALMA_observations}
    \resizebox{\hsize}{!}{%
    \begin{tabular}{cccclccc}
        \hline\hline
        Member OUS   & SB       & Array config. & Group & Galaxies & Synthesized Beam & CASA/Pipeline        \\ 
              &        &  &  &   & (\arcsec $\times$ \arcsec) &         \\ 
        (1)   & (2)       & (3) & (4) & (5) & (6) & (7)        \\ \hline

        uid://A001/X2f52/X2cf & 26668\_a\_03\_TM1 & C$-4$         & 1       & 26668, 27289, 27516, 27657, 46746  & 1.28 $\times$ 0.94 & 6.4.1.12/2022.2.0.64 \\
        uid://A001/X2f52/X2d4 & 55734\_a\_03\_TM1 & C$-4$         & 2       & 48125, 48131, 52854, 55734         & 1.23 $\times$ 0.86 & 6.5.4.9/2023.1.0.124 \\
        uid://A001/X2f52/X2d9 & 60871\_a\_03\_TM1 & C$-3$         & 3       & 59764, 59902, 60871                & 1.14 $\times$ 0.92 & 6.4.1.12/2022.2.0.64 \\
        uid://A001/X3577/X58d & 58741\_b\_03\_TM1 & C$-4$         & 4       & 58740, 58741, 58855                & 1.25 $\times$ 1.00 & 6.5.4.9/2023.1.0.124 \\
        uid://A001/X3577/X5e2 & 57508\_a\_03\_TM1 & C$-4$         & 4       & 57508                              & 1.29 $\times$ 0.96 & 6.5.4.9/2023.1.0.124 \\
        uid://A001/X2f52/X2e1 & 53609\_a\_03\_TM1 & C$-4$         & 5       & 50031, 53609, 54706                & 1.19 $\times$ 0.93 & 6.4.1.12/2022.2.0.64 \\
        uid://A001/X2f52/X2e6 & 65716\_a\_03\_TM1 & C$-3$         & 6       & 55131, 65288, 65716, 65783, 65887  & 1.09 $\times$ 0.97 & 6.4.1.12/2022.2.0.64 \\
        uid://A001/X2f52/X2e9 & 42443\_a\_03\_TM1 & C$-4$         & 7       & 42443                              & 1.85 $\times$ 0.82 & 6.4.1.12/2022.2.0.64 \\
        uid://A001/X2f52/X2ee & 37424\_a\_03\_TM1 & C$-3$         & 8       & 40294                              & 1.22 $\times$ 0.83 & 6.5.4.9/2023.1.0.124 \\
        uid://A001/X3578/Xe1  & 37424\_c\_03\_TM1 & C$-4$         & 8       & 37424, 42595, 49935, 51089         & 1.22 $\times$ 0.83 & 6.5.4.9/2023.1.0.124 \\
        uid://A001/X2f52/X2f3 & 34718\_a\_03\_TM1 & C$-3$         & 9       & 34718                              & 1.03 $\times$ 0.97 & 6.4.1.12/2022.2.0.64 \\
        uid://A001/X2f52/X2f6 & 38659\_a\_03\_TM1 & C$-3$         & 10      & 11248, 38659                       & 1.13 $\times$ 0.94 & 6.5.4.9/2023.1.0.124 \\
        uid://A001/X2f52/X2f9 & 32597\_a\_03\_TM1 & C$-3$         & 11      & 32522, 32597, 34439                & 1.19 $\times$ 0.87 & 6.4.1.12/2022.2.0.64 \\
        uid://A001/X2f52/X2fc & 63263\_a\_03\_TM1 & C$-3$         & 12      & 63263                              & 1.17 $\times$ 0.94 & 6.4.1.12/2022.2.0.64 \\
        uid://A001/X2f52/X2ff & 59266\_a\_03\_TM1 & C$-4$         & 13      & 53449, 58154, 59266, 62323         & 1.15 $\times$ 0.90 & 6.4.1.12/2022.2.0.64 \\ \hline
        uid://A001/X2f52/X2d1 & 26668\_a\_03\_TM2 & C$-1$         & 1       & 26668, 27289, 27516, 27657, 46746  & 3.60 $\times$ 2.71 & 6.5.4.9/2023.1.0.124 \\
        uid://A001/X2f52/X2d6 & 55734\_a\_03\_TM2 & C$-1$         & 2       & 48125, 48131, 52854, 55734         & 3.50 $\times$ 2.78 & 6.5.4.9/2023.1.0.124 \\
        uid://A001/X3577/X58f & 58741\_b\_03\_TM2 & C$-1$         & 4       & 58740, 58741, 58855                & 3.58 $\times$ 3.21 & 6.5.4.9/2023.1.0.124 \\
        uid://A001/X3577/X5e4 & 57508\_a\_03\_TM2 & C$-1$         & 4       & 57508                              & 3.47 $\times$ 2.56 & 6.5.4.9/2023.1.0.124 \\
        uid://A001/X2f52/X2e3 & 53609\_a\_03\_TM2 & C$-1$         & 5       & 50031, 53609, 54706                & 3.44 $\times$ 2.69 & 6.5.4.9/2023.1.0.124 \\
        uid://A001/X2f52/X2eb & 42443\_a\_03\_TM2 & C$-1$         & 7       & 42443                              & 3.51 $\times$ 2.81 & 6.5.4.9/2023.1.0.124 \\
        uid://A001/X2f52/X2f0 & 37424\_a\_03\_TM2 & C$-1$         & 8       & 40294                              & 3.88 $\times$ 2.68 & 6.5.4.9/2023.1.0.124 \\
        uid://A001/X3578/Xe3  & 37424\_c\_03\_TM2 & C$-1$         & 8       & 37424, 42595, 49935, 51089         & 3.58 $\times$ 2.74 & 6.5.4.9/2023.1.0.124 \\ 
        uid://A001/X2f52/X301 & 59266\_a\_03\_TM2 & C$-1$         & 13      & 53449, 58154, 59266, 62323         & 3.26 $\times$ 2.85 & 6.5.4.9/2023.1.0.124 \\ \hline

    \end{tabular}}

\tablefoot{(1) Member OUS: Member Observing Unit Set, (2) SB: Scheduling Block, (3) Array config.: ALMA array configurations, (4) Group: Group ID for galaxies in the different SBs, (5) Galaxies: CAVITY identifiers of the galaxies in each group, (6) Synthesised Beam: half power beam widths of the synthesised beam in arcsec, (7) CASA/Pipeline: CASA pipeline version used for the data processing. We separate datasets observed using C$-3$ and C$-4$ configurations from those observed with the C$-1$ configuration. }
    
\end{table*}

\begin{table*}[]
    \centering
    \caption{Properties of ALMA CO-CAVITY datacubes after data combination.}
    \label{tab:ALMA_products}
    \begin{tabular}{clclc}
        \hline\hline
        CAVITY ID & Array config.     & RMS  & Synthesised Beam   & Detection \\ 
                  &                   & ($\rm mJy\ beam^{-1}$) &    (\arcsec $\times$ \arcsec, \degrees)     \\ 
        (1)       & (2)               & (3)             & (4)          & (5) \\ \hline
        11248 & C$-3$ & 0.703 & $1.14 \times 0.92$, 89.0 & Detection \\
        26668 & C$-1$ + C$-4$ & 0.742 & $1.35 \times 1.01$, $-5.0$ & Tentative \\
        27289 & C$-1$ + C$-4$ & 0.727 & $1.40 \times 1.00$, $4.6$ & Detection \\
        27516 & C$-1$ + C$-4$ & 0.772 & $1.46 \times 1.01$, $5.9$ & Detection \\
        27657 & C$-1$ + C$-4$ & 0.688 & $1.35 \times 1.01$, $-3.0$ & Detection \\
        32522 & C$-3$ & 0.831 & $1.16 \times 0.88$, $-51.3$ & Non-detection \\
        32597 & C$-3$ & 0.833 & $1.20 \times 0.87$, $-50.4$ & Detection \\
        34439 & C$-3$ & 0.834 & $1.14 \times 0.88$, $-52.2$ & Non-detection \\
        34718 & C$-3$ & 0.936 & $1.04 \times 0.97$, $-44.6$ & Detection \\
        37424 & C$-1$ + C$-4$ & 0.897 & $1.28 \times 0.89$, $-12.3$ & Detection \\
        38659 & C$-3$ & 0.737 & $1.13 \times 0.95$, $-89.7$ & Detection \\
        40294 & C$-1$ + C$-3$ & 0.833 & $1.42 \times 1.01$, $6.5$ & Non-detection \\
        42443 & C$-1$ + C$-4$ & 0.837 & $1.89 \times 0.89$, $-39.3$ & Non-detection \\
        42595 & C$-1$ + C$-4$ & 0.880 & $1.30 \times 0.89$, $-14.4$ & Detection \\
        46746 & C$-1$ + C$-4$ & 0.736 & $1.35 \times 1.02$, $7.7$ & Non-detection \\
        48125 & C$-1$ + C$-4$ & 0.884 & $1.51 \times 1.06$, $-18.1$ & Detection \\
        48131 & C$-1$ + C$-4$ & 0.856 & $1.51 \times 1.06$, $-16.4$ & Non-detection \\
        49935 & C$-1$ + C$-4$ & 0.869 & $1.18 \times 0.93$, $-11.6$ & Tentative \\
        50031 & C$-1$ + C$-4$ & 0.757 & $1.39 \times 0.98$, $6.0$ & Detection \\
        51089 & C$-1$ + C$-4$ & 0.863 & $1.19 \times 0.91$, $-9.5$ & Detection \\
        52854 & C$-1$ + C$-4$ & 0.861 & $1.44 \times 1.08$, $-9.7$ & Detection \\
        53449 & C$-1$ + C$-4$ & 0.734 & $1.39 \times 0.98$, $-7.8$ & Detection \\
        53609 & C$-1$ + C$-4$ & 0.807 & $1.26 \times 1.01$, $7.4$ & Detection \\
        54706 & C$-1$ + C$-4$ & 0.773 & $1.24 \times 1.01$, $4.3$ & Detection \\
        55131 & C$-3$ & 0.909 & $1.14 \times 0.97$, $-20.9$ & Detection \\
        55734 & C$-1$ + C$-4$ & 0.874 & $1.44 \times 1.07$, $-9.5$ & Detection \\
        57508 & C$-1$ + C$-4$ & 0.817 & $1.39 \times 1.05$, $14.3$ & Detection \\
        58154 & C$-1$ + C$-4$ & 0.754 & $1.33 \times 0.98$, $-6.4$ & Detection \\
        58740 & C$-1$ + C$-4$ & 0.677 & $1.27 \times 1.05$, $-20.0$ & Detection \\
        58741 & C$-1$ + C$-4$ & 0.683 & $1.27 \times 1.05$, $-18.1$ & Detection \\
        58855 & C$-1$ + C$-4$ & 0.711 & $1.27 \times 1.04$, $-16.4$ & Detection \\
        59266 & C$-1$ + C$-4$ & 0.814 & $1.24 \times 0.99$, $-3.4$ & Detection \\
        59764 & C$-3$ & 0.794 & $1.22 \times 0.92$, $19.2$ & Detection \\
        59902 & C$-3$ & 0.815 & $1.19 \times 0.91$, $10.6$ & Detection \\
        60871 & C$-3$ & 0.834 & $1.15 \times 0.92$, $11.2$ & Detection \\
        62323 & C$-1$ + C$-4$ & 0.772 & $1.25 \times 0.98$, $-14.8$ & Detection \\
        63263 & C$-3$ & 1.040 & $1.17 \times 0.95$, $-18.2$ & Detection \\
        65288 & C$-3$ & 0.881 & $1.15 \times 0.97$, $-24.8$ & Detection \\
        65716 & C$-3$ & 0.912 & $1.09 \times 0.97$, $-34.8$ & Detection \\
        65783 & C$-3$ & 0.903 & $1.13 \times 0.97$, $-30.1$ & Detection \\
        65887 & C$-3$ & 0.890 & $1.14 \times 0.97$, $-23.7$ & Detection \\ \hline
    \end{tabular}

\tablefoot{
    (1) CAVITY ID: identifier in the CAVITY project, (2) Array config.: ALMA array configuration(s) used for the final ALMA data cubes, (3) RMS: root mean square of the ALMA data cubes in $\rm mJy\ beam^{-1}$ for a channel of 10 \kms, (4) Synthesised Beam: Half power beam widths of the synthesised beam, as well as its position angle (PA) measured from North to East. (5) Detection: detection label of the final ALMA data cubes. Detection is deemed when the S/N of the CO(1--0) integrated flux over the masked region is larger than 5, tentative detections 3 $<$ S/N $<$ 5 and non-detections otherwise. }

\end{table*}

\section{Potential flux loss due to lack of short spacings in ALMA observations}
\label{appendix:fluxloss}

The lack of short spacings in interferometric observations may lead to spatial filtering of extended emission.
In this appendix we compare the ALMA and IRAM 30m CO(1--0) integrated spectra and the derived fluxes, to quantify the potential loss, if any, in the interferometric observations. In \cref{fig:prof1,fig:prof2}, we observe that the spectra agree well in general, both the shape and the central velocities. 

\begin{figure*}
    \centering
    \includegraphics[width=\linewidth]{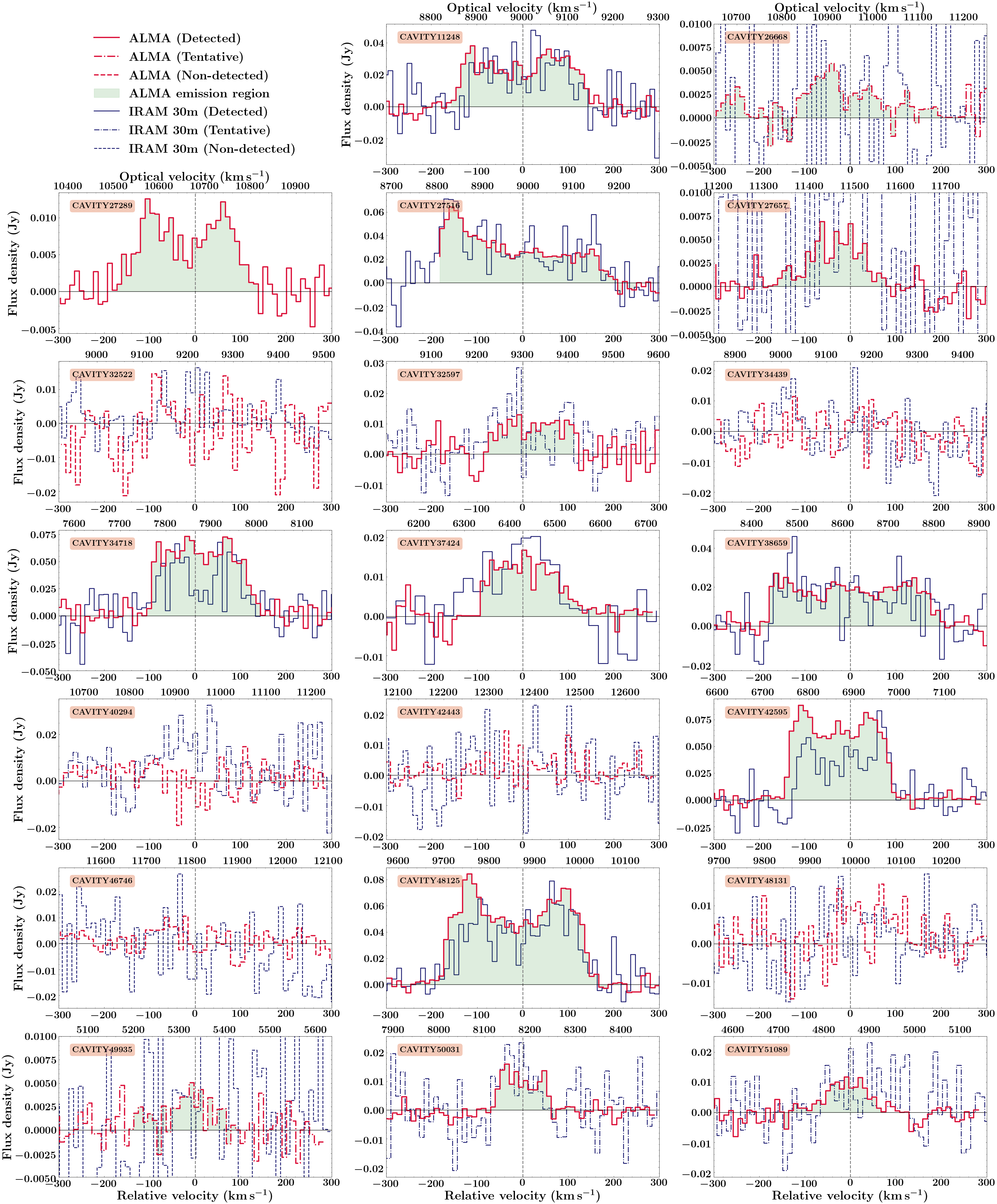}
    \caption{ALMA CO(1--0) spectra (red) for galaxies in ALMA CO-CAVITY. The velocities of all spectra are expressed in optical barycentric velocities. The vertical dashed lines are the velocities from SDSS ($cz$). IRAM 30m profiles from \citet{Rodriguez2024} are also shown for all galaxies in common for comparison. Full lines show detections, dot-dashed lines represent tentative detections, and dashed lines non-detections. Note that galaxies CAVITY IDs \texttt{27289}, \texttt{55734}, \texttt{57508}, and \texttt{59266} do not have IRAM 30m profiles.
    }
    \label{fig:prof1}
\end{figure*}

\begin{figure*}
    \centering
    \includegraphics[width=\linewidth]{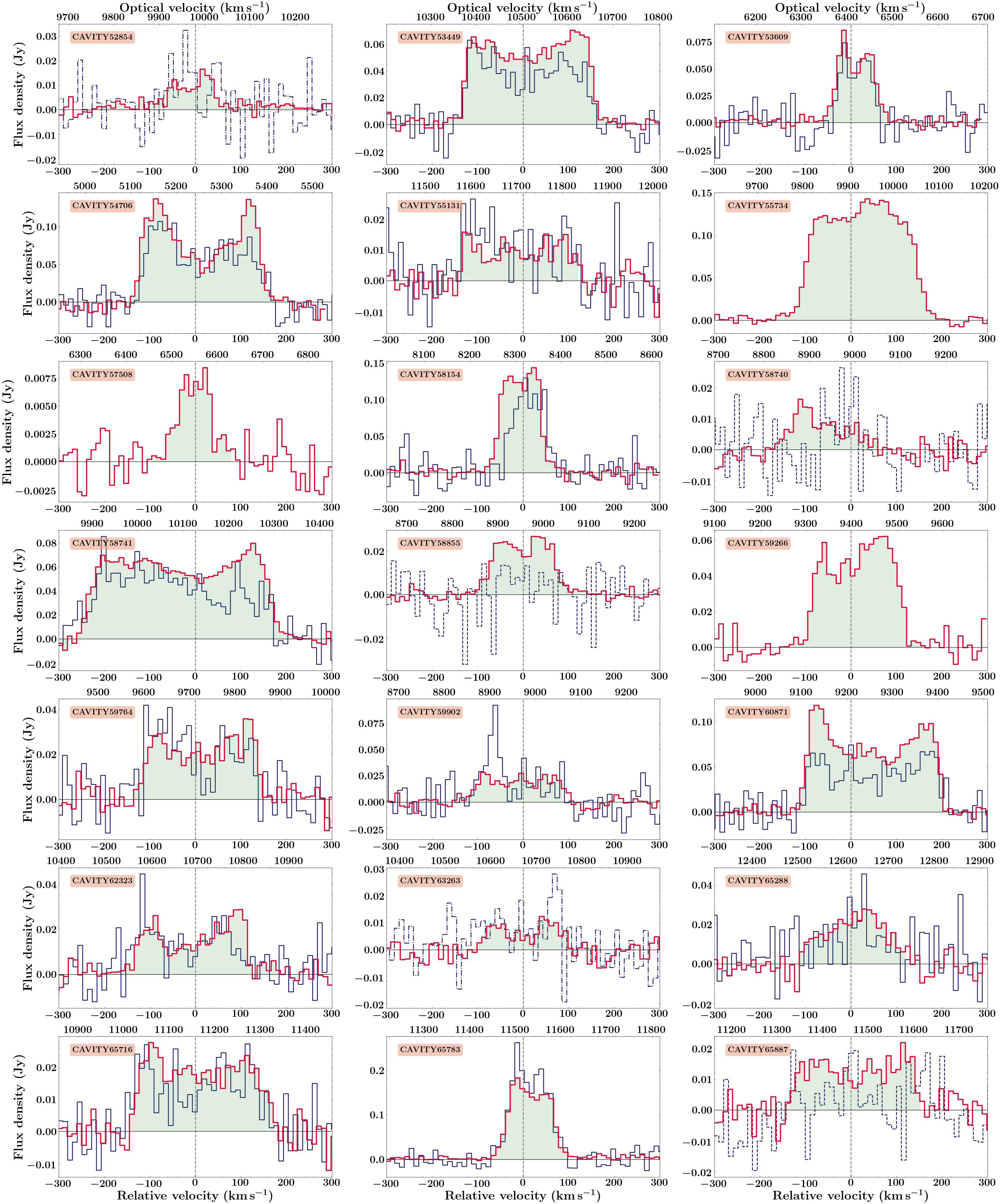}
    \caption{Continued from \cref{fig:prof1}.}
    \label{fig:prof2}
\end{figure*}

\begin{figure}
    \centering
    \includegraphics[width=0.6\linewidth]{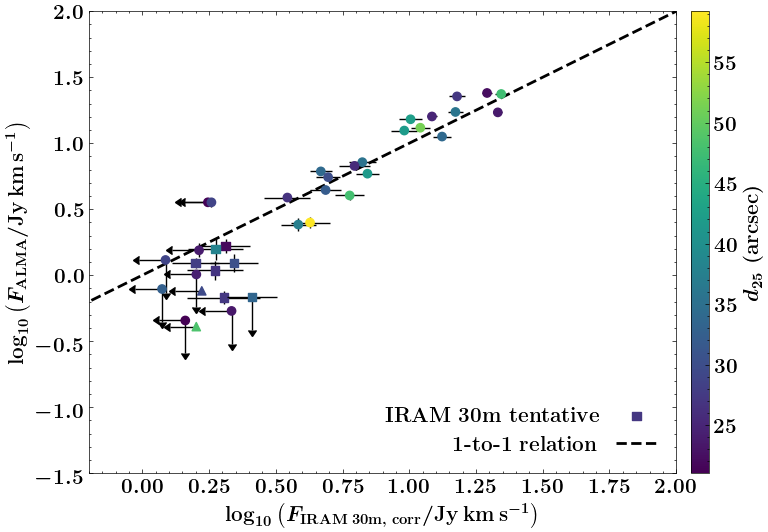}
    \caption{Comparison between ALMA ($F_{\rm ALMA}$) and IRAM 30m CO(1--0) fluxes (aperture corrected, $F_{\rm IRAM~30m,~corr}$) (in decimal log scale) for the 37 galaxies that overlap between ALMA CO-CAVITY and CO-CAVITY. Arrows show CO(1--0) flux upper limits in either data product of each instrument. The dashed line is the 1-to-1 relation. Tentative detections from the IRAM 30m observations are depicted with square symbols as in \citet{Rodriguez2024}. The colour scale of the data points shows the diameter $d_{25}$.}
    \label{fig:compareflux}
\end{figure}

To obtain the IRAM 30m CO(1--0) fluxes, a K to Jy conversion factor of 5 has been applied \citep[e.g.][]{Rodriguez2024}. Note that we have applied an aperture correction as in \citet{Rodriguez2024} to the IRAM 30m CO(1--0) fluxes, which depends on the galaxy size but account typically to no more than 10\%. The velocity-integrated fluxes obtained from the ALMA and IRAM 30m CO(1--0) profiles are consistent to each other and they are usually within their uncertainties (see \cref{fig:compareflux} and \cref{tab:molecular}), especially at the high CO(1--0) flux end. The median fraction of IRAM fluxes with respect to ALMA fluxes for tentative and detected galaxies is 1.02, with a median absolute deviation of 0.22, which confirms that almost no flux was lost as a result of missing short-spacings in the ALMA observations compared to the single-dish data from IRAM 30m telescope. Nonetheless, there are measurements in disagreement, in particular in the low-flux regime, where only tentative detections or non-detections are achieved in one of the telescopes. We note that there are cases where estimated fluxes do not seem compatible, but given the higher sensitivity of ALMA, and the emission size per channel compared to the Largest Angular Scale (LAS) of the observations, they are not dominated by the missing zero-spacing problem. 
After reviewing IRAM 30m spectra, the fluxes might be affected by baseline problems that may increase uncertainty, yielding in a few cases tentative or possibly false detections.

\begin{table*}[]
    \centering
    \caption{Global molecular gas properties of the ALMA CO-CAVITY sample.}
    \begin{tabular}{cccccc}
        \hline\hline
        CAVITY ID  & $F_{\rm IRAM~30m,~corr}$ & $F_{\rm ALMA }$ & $\alpha_{\rm CO}$ & $\log(M_{\rm H_2,\ 30m})$ & $\log(M_{\rm H_2,\ ALMA})$  \\
                   & (Jy \kms)      & (Jy \kms)    & ($\rm M_\odot\ (K\ km\ s^{-1}\ pc^2)^{-1}$)  & log([$\rm M_\odot$])   & log([$\rm M_\odot$])\\
                   (1)  & (2)   & (3)  & (4)        & (5)           & (6)\\ \hline
        11248 & $6.7 \pm 0.8$ & $7.2 \pm 0.3$ & 3.13 & $8.93 \pm 0.06$ & $8.966 \pm 0.017$ \\
        26668 & $<1.67$ & $0.76 \pm 0.17$ & 5.82 & $9.13 \pm 0.14$ & $8.43 \pm 0.10$ \\
        27289 &  & $1.98 \pm 0.13$ & 4.18 &  & $8.68 \pm 0.03$ \\
        27516 & $13.3 \pm 1.1$ & $11.2 \pm 0.3$ & 3.18 & $9.24 \pm 0.05$ & $9.166 \pm 0.011$ \\
        27657 & $2.0 \pm 0.6$ & $0.68 \pm 0.08$ & 8.45 & $9.06 \pm 0.15$ & $8.59 \pm 0.05$ \\
        32522 & $<1.22$ & $<1.30$ & 6.37 & $<8.51$ & $<8.55$ \\
        32597 & $2.1 \pm 0.4$ & $1.67 \pm 0.21$ & 5.76 & $8.71 \pm 0.10$ & $8.62 \pm 0.05$ \\
        34439 & $<1.59$ & $<1.01$ & 6.03 & $<8.60$ & $<8.41$ \\
        34718 & $9.6 \pm 1.1$ & $12.5 \pm 0.5$ & 3.16 & $8.98 \pm 0.06$ & $9.092 \pm 0.017$ \\
        37424 & $3.8 \pm 0.6$ & $2.4 \pm 0.3$ & 4.04 & $8.5 \pm 0.3$ & $8.31 \pm 0.05$ \\
        ... \\ \hline
    \end{tabular}
    \tablefoot{
    (This table is truncated to the first 10 entries. The complete table is presented in its entirety in the online version of the publication.) 
    (1) ID: CAVITY identifier in the CAVITY sample. 
    (2) $F_{\rm IRAM~30m,~corr}$: The aperture-corrected flux obtained from CO-CAVITY IRAM 30m observations, compiled from \citet{Rodriguez2024}. Note that galaxies with blank values were not observed.
    (3) $F_{\rm ALMA}$: The flux obtained from ALMA CO-CAVITY observations following the procedure in \cref{sec:alma}.
    (4) $\alpha_{\rm CO}$: CO conversion factor, as in \cite{Rodriguez2024}. 
    (5) (6) $\log(M_{\rm H_2,\ ALMA})$) and $\log(M_{\rm H_2,\ 30m})$: molecular gas mass in decimal log scale from ALMA CO-CAVITY and \cite{Rodriguez2024} using $\alpha_{\rm CO}$ in column (4), respectively. 
    }

    \label{tab:molecular}
\end{table*}

To estimate the uncertainties on the integrated fluxes derived from the CO(1--0) profiles, we used the expression $\sigma \times \sqrt{\delta v\ \Delta V}$, where $\sigma$ is the RMS measured in emission-free regions of the integrated spectrum for each galaxy, $\delta v$ is the velocity width of a channel ($\sim 10\ {\rm km\ s^{-1}}$), and $\Delta V$ is the velocity range over which the line emission extends. For non-detections, we assume a conservative value of $\Delta V = 300\ {\rm km\ s^{-1}}$.

\end{document}